\pdfoutput=1
\documentclass[12pt,a4paper]{article}
\usepackage{comment}
\usepackage{ifthen} 
\usepackage{comment}
\newboolean{pdflatex}
\setboolean{pdflatex}{true} 

\newboolean{articletitles}
\setboolean{articletitles}{true} 

\newboolean{uprightparticles}
\setboolean{uprightparticles}{false} 

\def\paperauthors{LHCb collaboration} 
\def\paperasciititle{SemiCharmTag: a tool for Semileptonic Charm tagging} 
\def\papertitle{SemiCharmTag: a tool for Semileptonic Charm tagging} 
\def\paperkeywords{{High Energy Physics}, {LHCb}} 
\def\papercopyright{\the\year\ CERN for the benefit of the LHCb collaboration} 
\def\paperlicence{CC BY 4.0 licence}
\def\paperlicenceurl{https://creativecommons.org/licenses/by/4.0/}

\newif\ifEnableSectionTOCLinks
\EnableSectionTOCLinksfalse 

\usepackage[top=1in, bottom=1.25in, left=1in, right=1in]{geometry}

\usepackage{ifthen}
\usepackage{microtype}
\usepackage{lineno}  
\usepackage{xspace} 
\usepackage{caption} 
\usepackage{subcaption}

\usepackage{graphicx}  
\usepackage{color}
\usepackage{colortbl}

\usepackage{amsmath} 
\usepackage{amssymb}
\usepackage{amsfonts}
\usepackage{upgreek} 

\newcommand*\patchAmsMathEnvironmentForLineno[1]{%
\expandafter\let\csname old#1\expandafter\endcsname\csname #1\endcsname
\expandafter\let\csname oldend#1\expandafter\endcsname\csname
end#1\endcsname
 \renewenvironment{#1}%
   {\linenomath\csname old#1\endcsname}%
   {\csname oldend#1\endcsname\endlinenomath}%
}
\newcommand*\patchBothAmsMathEnvironmentsForLineno[1]{%
  \patchAmsMathEnvironmentForLineno{#1}%
  \patchAmsMathEnvironmentForLineno{#1*}%
}
\AtBeginDocument{%
\patchBothAmsMathEnvironmentsForLineno{equation}%
\patchBothAmsMathEnvironmentsForLineno{align}%
\patchBothAmsMathEnvironmentsForLineno{flalign}%
\patchBothAmsMathEnvironmentsForLineno{alignat}%
\patchBothAmsMathEnvironmentsForLineno{gather}%
\patchBothAmsMathEnvironmentsForLineno{multline}%
\patchBothAmsMathEnvironmentsForLineno{eqnarray}%
}

\usepackage[pdftex,
            pdfauthor={\paperauthors},
            pdftitle={\paperasciititle},
            pdfkeywords={\paperkeywords}]{hyperref}
\usepackage{hyperxmp}
\hypersetup{
    pdfcopyright={Copyright (C) \papercopyright},
    pdflicenseurl={\paperlicenceurl}
}

\usepackage[colorinlistoftodos,textsize=scriptsize]{todonotes}

\usepackage[bottom,flushmargin,hang,multiple]{footmisc}

\usepackage[all]{hypcap} 

\usepackage{xspace}
\usepackage{upgreek}

\def\lhcb   {\mbox{LHCb}\xspace}

\def\MagUp {\mbox{\em Mag\kern -0.05em Up}\xspace}

\ifthenelse{\boolean{uprightparticles}}%
{

 \def\PDelta      {\ensuremath{\Delta}\xspace}
 \def\PXi         {\ensuremath{\Xi}\xspace}
 \def\PLambda     {\ensuremath{\Lambda}\xspace}
 \def\PSigma      {\ensuremath{\Sigma}\xspace}
 \def\POmega      {\ensuremath{\Omega}\xspace}
 \def\PUpsilon    {\ensuremath{\Upsilon}\xspace}
 \let\oldPi\Pi
 \def\PPi         {\ensuremath{\oldPi}\xspace}

 \def\PB      {\ensuremath{\mathrm{B}}\xspace}
 \def\PD      {\ensuremath{\mathrm{D}}\xspace}

 \def\PK      {\ensuremath{\mathrm{K}}\xspace}

 \def\Pc      {\ensuremath{\mathrm{c}}\xspace}

 \def\Ph      {\ensuremath{\mathrm{h}}\xspace}
 \def\Ps      {\ensuremath{\mathrm{s}}\xspace}

 \def\thebaroffset{0.0em}
}
{

 \mathchardef\PDelta="7101
 \mathchardef\PXi="7104
 \mathchardef\PLambda="7103
 \mathchardef\PSigma="7106
 \mathchardef\POmega="710A
 \mathchardef\PUpsilon="7107
 \mathchardef\PPi="7105
 \def\PB      {\ensuremath{B}\xspace}
 \def\PD      {\ensuremath{D}\xspace}

 \def\PK      {\ensuremath{K}\xspace}

 \def\Pc      {\ensuremath{c}\xspace}

 \def\Ph      {\ensuremath{h}\xspace}
 \def\Ps      {\ensuremath{s}\xspace}

 \def\thebaroffset{0.18em}
}
\newcommand{\offsetoverline}[2][\thebaroffset]{\kern #1\overline{\kern -#1 #2}}%

\makeatletter
\ifcase \@ptsize \relax
  \newcommand{\miniscule}{\@setfontsize\miniscule{4}{5}}
\or
  \newcommand{\miniscule}{\@setfontsize\miniscule{5}{6}}
\or
  \newcommand{\miniscule}{\@setfontsize\miniscule{5}{6}}
\fi
\makeatother

\DeclareRobustCommand{\optbar}[1]{\shortstack{{\miniscule (\rule[.5ex]{1.25em}{.18mm})}
  \\ [-.7ex] $#1$}}

\def\squark    {{\ensuremath{\Ps}}\xspace}

\def\cquark    {{\ensuremath{\Pc}}\xspace}

\def\hadron {{\ensuremath{\Ph}}\xspace}

\def\KorKbar {\kern \thebaroffset\optbar{\kern -\thebaroffset \PK}{}\xspace}

\def\D       {{\ensuremath{\PD}}\xspace}

\def\DorDbar {\kern \thebaroffset\optbar{\kern -\thebaroffset \PD}\xspace}
\def\Dz      {{\ensuremath{\D^0}}\xspace}

\def\Dp      {{\ensuremath{\D^+}}\xspace}
\def\Dm      {{\ensuremath{\D^-}}\xspace}

\def\DpDm    {\ensuremath{\Dp {\kern -0.16em \Dm}}\xspace}

\def\Dsp     {{\ensuremath{\D^+_\squark}}\xspace}

\def\B       {{\ensuremath{\PB}}\xspace}

\def\BorBbar {\kern \thebaroffset\optbar{\kern -\thebaroffset \PB}\xspace}

\def\Bd      {{\ensuremath{\B^0}}\xspace}

\def\BdorBdbar {\kern \thebaroffset\optbar{\kern -\thebaroffset \Bd}\xspace}

\def\Bs      {{\ensuremath{\B^0_\squark}}\xspace}

\def\BsorBsbar {\kern \thebaroffset\optbar{\kern -\thebaroffset \Bs}\xspace}

\def\Y#1S{\ensuremath{\PUpsilon{(#1S)}}\xspace}

\def\Lz          {{\ensuremath{\PLambda}}\xspace}

\def\LorLbar     {\kern \thebaroffset\optbar{\kern -\thebaroffset \PLambda}\xspace}

\def\Xires       {{\ensuremath{\PXi}}\xspace}

\def\Omegares    {{\ensuremath{\POmega}}\xspace}

\def\Lc          {{\ensuremath{\Lz^+_\cquark}}\xspace}

\def\Xic         {{\ensuremath{\Xires_\cquark}}\xspace}

\def\Xicp        {{\ensuremath{\Xires^+_\cquark}}\xspace}

\def\Omegac      {{\ensuremath{\Omegares^0_\cquark}}\xspace}

\def\to                 {\ensuremath{\rightarrow}\xspace}

\def\AT#1     {\ensuremath{A_{\mathrm{T}}^{#1}}\xspace}           

\def\C#1      {\ensuremath{\mathcal{C}_{#1}}\xspace}                       
\def\Cp#1     {\ensuremath{\mathcal{C}_{#1}^{'}}\xspace}                    
\def\Ceff#1   {\ensuremath{\mathcal{C}_{#1}^{\mathrm{(eff)}}}\xspace}        
\def\Cpeff#1  {\ensuremath{\mathcal{C}_{#1}^{'\mathrm{(eff)}}}\xspace}       
\def\Ope#1    {\ensuremath{\mathcal{O}_{#1}}\xspace}                       
\def\Opep#1   {\ensuremath{\mathcal{O}_{#1}^{'}}\xspace}                    

\newcommand{\nospaceunit}[1]{\ensuremath{\text{#1}}}
\newcommand{\aunit}[1]{\ensuremath{\text{\,#1}}}

\newcommand{\tev}{\aunit{Te\kern -0.1em V}\xspace}
\newcommand{\gev}{\aunit{Ge\kern -0.1em V}\xspace}
\newcommand{\mev}{\aunit{Me\kern -0.1em V}\xspace}
\newcommand{\kev}{\aunit{ke\kern -0.1em V}\xspace}
\newcommand{\ev}{\aunit{e\kern -0.1em V}\xspace}

\newcommand{\mevc}{\ensuremath{\aunit{Me\kern -0.1em V\!/}c}\xspace}
\newcommand{\gevc}{\ensuremath{\aunit{Ge\kern -0.1em V\!/}c}\xspace}
\newcommand{\mevcc}{\ensuremath{\aunit{Me\kern -0.1em V\!/}c^2}\xspace}
\newcommand{\gevcc}{\ensuremath{\aunit{Ge\kern -0.1em V\!/}c^2}\xspace}

\def\mm   {\aunit{mm}\xspace}

\def\mum  {\ensuremath{\,\upmu\nospaceunit{m}}\xspace}

\newcommand{\chisq}{\ensuremath{\chi^2}\xspace}

\newcommand{\chisqip}{\ensuremath{\chi^2_{\text{IP}}}\xspace}
\newcommand{\chisqfd}{\ensuremath{\chi^2_{\text{FD}}}\xspace}

\newcommand{\chisqvtx}{\ensuremath{\chi^2_{\text{vtx}}}\xspace}

\def\gsim{{~\raise.15em\hbox{$>$}\kern-.85em
          \lower.35em\hbox{$\sim$}~}\xspace}
\def\lsim{{~\raise.15em\hbox{$<$}\kern-.85em
          \lower.35em\hbox{$\sim$}~}\xspace}

\def\sqs   {\ensuremath{\protect\sqrt{s}}\xspace}

\def\pt         {\ensuremath{p_{\mathrm{T}}}\xspace}

\def\ptot       {\ensuremath{p}\xspace}

\def\davinci    {\mbox{\textsc{DaVinci}}\xspace}

\def\evtgen     {\mbox{\textsc{EvtGen}}\xspace}

\def\gaudi      {\mbox{\textsc{Gaudi}}\xspace}
\def\gauss      {\mbox{\textsc{Gauss}}\xspace}
\def\geant      {\mbox{\textsc{Geant4}}\xspace}

\def\photos     {\mbox{\textsc{Photos}}\xspace}

\def\pythia     {\mbox{\textsc{Pythia}}\xspace}

\def\tell1  {TELL1\xspace}
\def\ukl1   {UKL1\xspace}

\newcommand{\lhcborcid}[1]{\href{https://orcid.org/#1}{\hspace*{0.1em}\raisebox{-0.45ex}{\includegraphics[width=1em]{figs/orcidIcon.pdf}}}}

\hypersetup{
  colorlinks   = true, 
  urlcolor     = blue, 
  linkcolor    = blue, 
  citecolor    = red   
}

\usepackage{cite} 
\usepackage{mciteplus}

\usepackage{longtable} 
\usepackage{subcaption}

\begin{document}

\renewcommand{\thefootnote}{\fnsymbol{footnote}}
\setcounter{footnote}{1}


\begin{titlepage}
\pagenumbering{roman}

\vspace*{-1.5cm}
\centerline{\large EUROPEAN ORGANIZATION FOR NUCLEAR RESEARCH (CERN)}
\vspace*{1.5cm}
\noindent
\begin{tabular*}{\linewidth}{lc@{\extracolsep{\fill}}r@{\extracolsep{0pt}}}
\ifthenelse{\boolean{pdflatex}}
{\vspace*{-1.5cm}\mbox{\!\!\!\includegraphics[width=.14\textwidth]{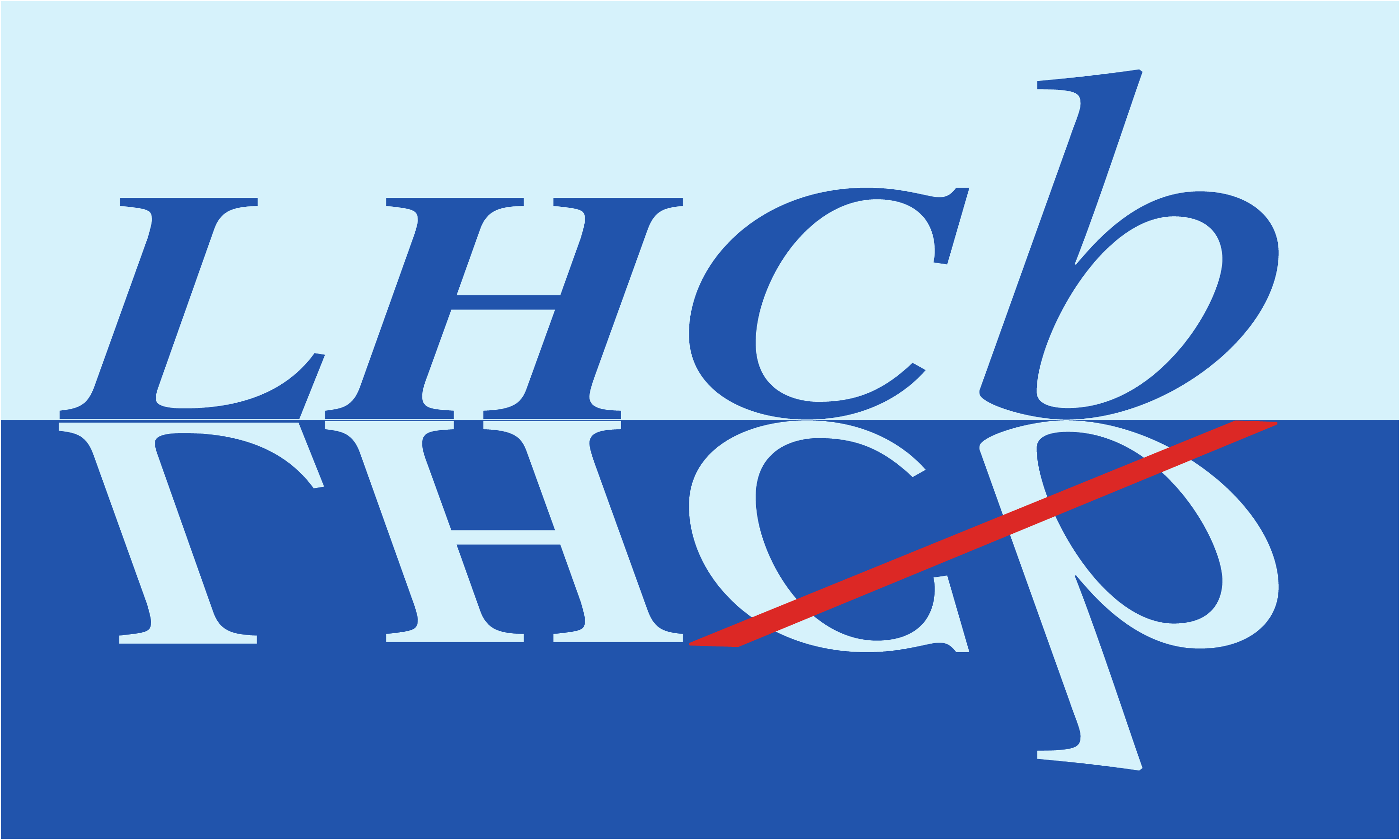}} & &}%
{\vspace*{-1.2cm}\mbox{\!\!\!\includegraphics[width=.12\textwidth]{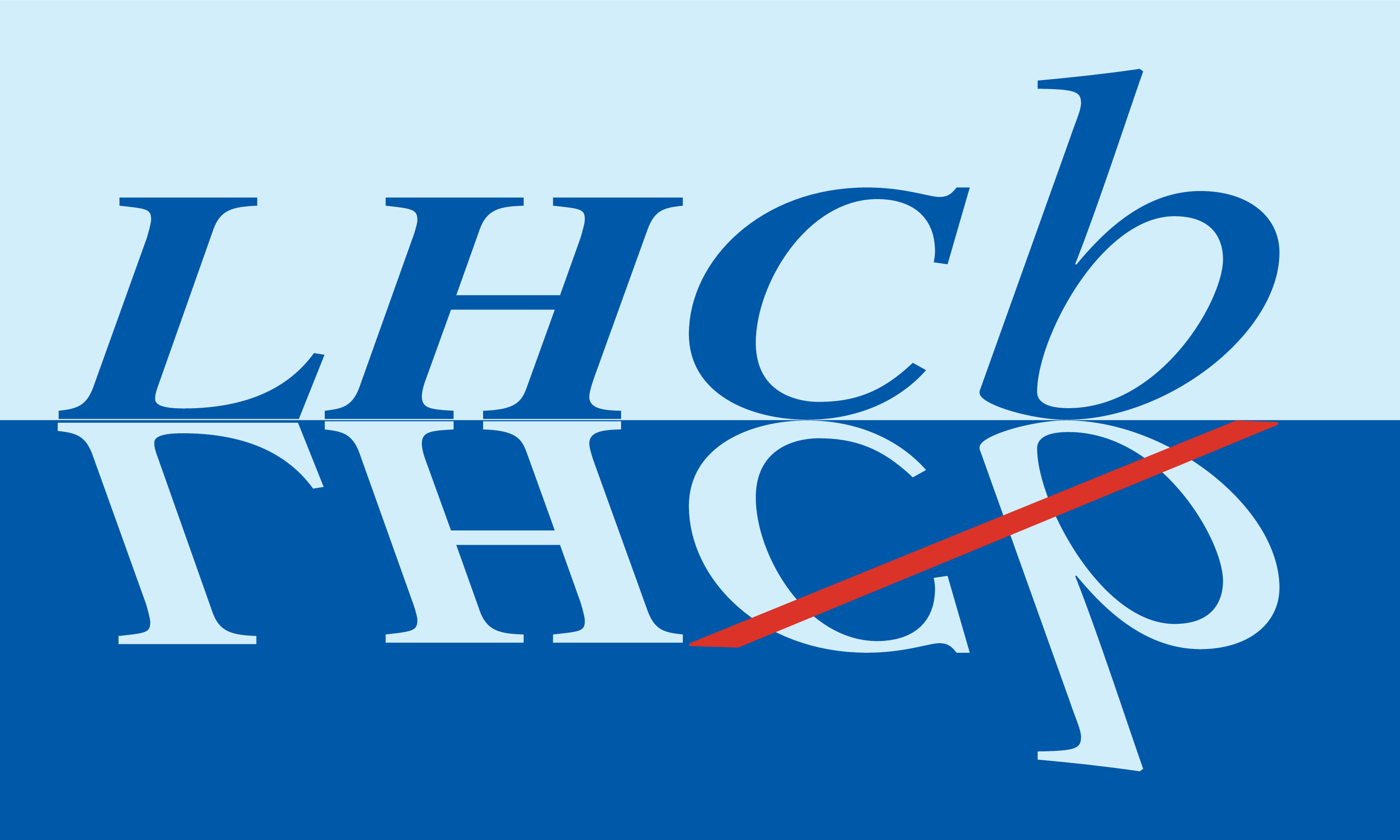}} & &}%
\\
 & & CERN-LHCb-RD-2026-001 \\  
 & & \today \\ 
 & & \\
\end{tabular*}

\vspace*{4.0cm}

{\normalfont\bfseries\boldmath\huge\begin{center}
  \papertitle
\end{center}
}

\vspace*{2.0cm}

\begin{center}
C. Arata$^{1}$, I. Corredoira$^{1}$, A. Lightbody$^{1}$, M. Winn$^{1}$\\
\vspace{0.2cm}
\textit{\small$^{1}$ Université Paris-Saclay, Centre d’Études de Saclay (CEA), IRFU, Gif-sur-Yvette, France}
\end{center}

\vspace{\fill}

\begin{abstract}
  \noindent
A method for selecting and/or rejecting leptons from charm semileptonic decays based on the tagging of the secondary vertex using a hadron track is introduced. 
The method is developed for dimuon Drell-Yan measurements in LHCb using full simulations in proton-proton collisions at $\sqrt{s}=13.6$~\tev. We focus on the invariant mass range between 2.9 and 5 \gevcc with single muon transverse momentum larger than 1 GeV/$c$. 
A novel strategy is detailed for background rejection, achieving an improvement of the signal over background of a factor $\sim 4$ at an efficiency of $81\%$ while maintaining the Drell-Yan kinematic distributions largely unbiased except at the acceptance edges. Moreover, a second approach is presented for the construction of unbiased background-pure samples of single muons from charm decays, achieving a charm efficiency of 21.4\% at a Drell-Yan efficiency of 1.1\%. 
  
\end{abstract}

\vspace*{2.0cm}

\begin{center}
  To be submitted to Journal of Physics G
\end{center}

\vspace{\fill}

{\footnotesize 
\centerline{\copyright~\papercopyright. \href{\paperlicenceurl}{\paperlicence}.}}
\vspace*{2mm}

\end{titlepage}


\newpage
\setcounter{page}{2}
\mbox{~}
%
%
%
%


\renewcommand{\thefootnote}{\arabic{footnote}}
\setcounter{footnote}{0}


\cleardoublepage


\pagestyle{plain} 
\setcounter{page}{1}
\pagenumbering{arabic}


\clearpage
\section{Introduction}
\label{sec:Introduction}

The production of dileptons in hadron-hadron collisions enables us to infer properties of the emitting source. In proton-proton or proton-nucleus collisions, the dilepton production via the Drell-Yan process is used to constrain the collinear parton distribution functions (PDFs) of the colliding hadrons~\cite{Ethier:2020way, Klasen:2023uqj} as well as the transverse momentum dependent distributions (TMDs)~\cite{Boussarie:2023izj}. In nucleus-nucleus collisions, thermal and preequilibrium production of dileptons contains information on the space-time evolution and the properties of the produced strongly interacting matter~\cite{Salabura:2020tou, Coquet:2021lca,Bailhache:2025kwa}.

Measurements in the lowest possible dilepton-mass interval amenable to perturbative quantum chromodynamics~(pQCD), namely 2-10 GeV/$c^2$, are of particular relevance. They provide a sensitive test of the applicability of pQCD, push the kinematic reach of the partonic system down to the lowest possible longitudinal momentum fractions interesting for the search of gluon saturation~\cite{Gelis:2010nm}, and offer access to thermal~\cite{Bailhache:2025kwa} and preequilibrium dilepton~\cite{Coquet:2021lca} production. In this kinematic domain, measurements at centre-of-mass energies of 100 \gev and above are particularly challenging due to the presence of a large background. This background consists of dileptons from semileptonic weak decays of charm- and beauty-hadron pairs produced by the strong interaction.

The signal extraction in Drell-Yan, thermal, or preequilibrium dilepton production measurements is based on identifying distributions in one or several discriminative variables between signal and background candidates. The distinction between the background from heavy-flavour semileptonic decays and the dilepton signal can be achieved by observables sensitive to the fact that the background leptons stem from the heavy-flavour hadron decay vertex, whereas the signal leptons originate from the primary vertex. The signal extraction will be based on fitting the data distribution with different functional forms for the signal and background contributions. The normalization of the signal contribution corresponds to the raw yield for the cross-section measurement. In the following, we assume this type of signal extraction strategy for dilepton measurements. Given the small signal over background ratio, the feasibility of this approach relies on the accuracy of the distributions used for the signal and background contributions. The distribution of the discriminative observables for the prompt dileptons can be extracted from data or the simulation can be tuned based on the large statistics from prompt dilepton resonances $\phi, \text{J}/\psi, \Upsilon (1S)$, where the non-prompt J/$\psi$ must be dealt with. The distribution for the contribution from beauty hadron decays is assumed to be obtained from the simulation.  

In this publication, a strategy to construct unbiased background distributions from data for semileptonic decays of charm hadrons and a strategy to improve the signal over background ratio in dimuon measurements are presented. In Sec.~\ref{sec:mot}, the motivation is explained. Afterwards, a description of the kinematic domain, the simulation datasets, and the selection criteria used in the article are detailed in Sec.~\ref{sec:selection}. It follows an estimation of the signal over background ratio based on NLO calculations in Sec.~\ref{sec:soverbest}. After the introduction of the algorithm and its implementation in Sec.~\ref{sec:alg}, the double-tag strategy for signal-over-background improvements is presented in Sec.~\ref{sec:double} and the single-tag strategy for background template construction is presented in Sec.~\ref{sec:single}. Finally,  a summary and outlook are provided in Sec.~\ref{sec:out}. 

\section{Motivation for SemiCharmTag}
\label{sec:mot}
The background from semileptonic decays of charm and beauty hadrons is, in principle, reducible since these leptons stem from weak decays of hadrons with lifetimes in the range $c \tau \approx 50-500$\mum: at finite momentum of the charm or beauty hadron, the impact parameter (IP), defined as the closest distance between the interaction point and the reconstructed lepton track ~\cite{LHCb-TDR-013}, will exhibit a distribution different from the one of Drell-Yan, thermal or preequilibrium leptons, which originate directly from the primary vertex. In the following, we will use the notion of prompt leptons (Drell-Yan, thermal, preequilibrium) and non-prompt leptons (weak decays of charm and beauty). 

The inclusive charm production dominates the beauty production over the full phase space. The dilepton production from charm can be up to 100 times larger than the Drell-Yan production at the LHC. The longer lifetimes of beauty hadrons facilitate the discrimination with respect to prompt dileptons compared to charm hadrons. The charm background is, therefore, more challenging to deal with. 


With the advent of modern silicon vertex detectors, the separation of prompt and non-prompt dileptons has been achieved in past measurements using the IP or related observables ~\cite{NA60:2008ctj, CMS:2021ynu, CDF_Jpsi,ALICE:2018fvj,ALICE:2023jef}, whose resolution can be well modeled in simulation. In addition, heavy quarkonium resonance production, in particular the $\Upsilon$(1S) as well as J/$\psi$ not originating from b-hadron decays, can serve as calibration candles for prompt dileptons. 

Performing the measurement in a fiducial region with small transverse momenta (\pt)  presents several challenges. In particular, the use of an isolation criterium has been pursued (e.g. in~\cite{ATLAS:2014ape,CMS:2021ynu}); however, its application  becomes increasingly challenging at low invariant masses and for lepton transverse momenta below \pt$<10$\gevc. Decreasing mass and transverse momentum scales lead to an increasing sensitivity to higher order corrections, to the modelling of the parton shower, hadronisation and the underlying event that are not well modelled in simulation. Ref.~\cite{ALICE:2019wqv} illustrates large data-simulation discrepancies at jet transverse momenta below 10\gevc. These discrepancies between simulation and data could be overcome with data-driven calibration. However, a data-driven approach with control samples in data is not achievable, as heavy-flavour decays dominate the dilepton spectra at low transverse momentum and mass scales, and there is no pure source of Drell-Yan, thermal or preequilibrium dileptons in this region. 

\begin{figure}[htb!]
    \centering
    \includegraphics[width=0.7\linewidth]{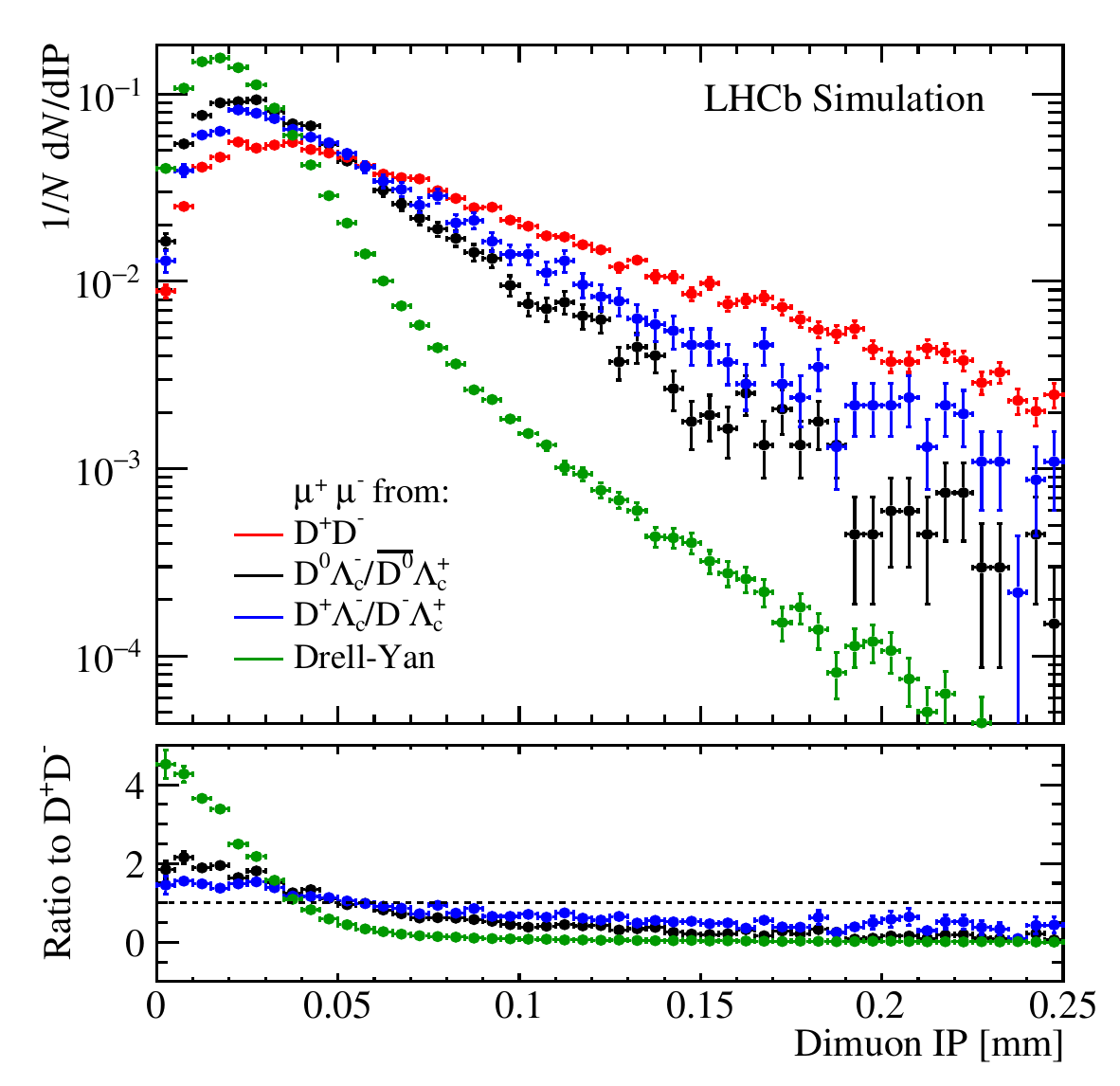}
    \caption{Top: dimuon IP distributions for dimuons coming from \Dp\Dm  (red-full circles), \Dz$\rm{\Lambda_{c}^{-}}$ ($\overline{\rm{D^{0}}}$\Lc) (black-full circles),  \Dp$\rm{\Lambda_{c}^{-}}$ (\Dm\Lc) (blue-full circles) and Drell-Yan (green-full circles) in full simulation. Bottom: ratio of dimuon IP distributions for Drell-Yan, \Dz${\Lambda_{c}^{-}}$ ($\overline{{D^{0}}}$\Lc) and \Dp${\Lambda_{c}^{-}}$ (\Dm$\Lambda_{c}^{+}$) with respect to dimuons from \Dp\Dm.}
    \label{fig:displacement}
\end{figure}

In addition to the initially small signal over background ratio, the precise measurement of prompt dileptons in the invariant mass region 2-30\gevcc down to low dilepton transverse momentum is complicated by the limited knowledge on the fragmentation fractions of different charm hadrons~\cite{Apolinario:2022vzg}. This is relevant since the different charm hadrons exhibit large decay length differences, ranging from $c\tau =45\mum$ to $c\tau =310\mum$ ~\cite{PDG2024} as it can be seen in Fig.~\ref{fig:displacement} and detailed in the following. 
In particular, the fraction of baryons is found to be more important in proton-proton collisions at midrapidity~\cite{ALICE:2021dhb,ALICE:2023sgl} than at the Large Electron-Positron Collider (LEP), where precise measurements are available~\cite{ALEPH:2005ab}. In particular, the fraction of the $\Lambda_c^+$ baryon has been found to be about a factor 3 more important~\cite{ALICE:2021dhb,ALICE:2023sgl} at central rapidity at the LHC than in event generator estimates based on jet universality between $e^+e^-$ collisions and hadron-hadron collisions~\cite{Skands:2014pea,Sjostrand:2014zea,Bellm:2015jjp}. At forward rapidity, an early $\Lambda_c^+$ baryon production cross section measurement does not indicate this stark difference~\cite{LHCb:2013xam}. A comprehensive set of measurements of all weakly decaying baryons at different centre-of-mass energies, collision systems and rapidities is still missing by ALICE and \lhcb. 

Moreover, the inclusive semileptonic decay branching fractions of charm baryons are unknown to a large extent~\cite{PDG2024}. The ongoing measurement programs at the electron-positron collider facilities BESIII and Belle2  may improve the situation in the near future~\cite{Li:2025nzx}. 
The impact of this missing knowledge on the non-prompt dileptons can be illustrated by using observables that separate prompt and non-prompt dileptons, such as the IP. Fig.~\ref{fig:displacement} shows that the precise knowledge of the baryon fractions has a crucial impact on the shapes of the IP distributions used for this separation. 

In the absence of sufficiently precise measurements of charm production and charm decays, data-driven methods are mandatory to avoid a model dependence of measurement cross sections. In a low signal over background situation, they may be required to render a measurement possible. The data-driven methods must target to quantify in data the relevant characteristics of semileptonic charm decays. Furthermore, any controlled sizable improvement of the signal over background using the characteristics of weak decays of charm will lead to immediate improvements of the systematic uncertainties.

The $\mathtt{SemiCharmTag}$ approach presented in the following is the first method to derive data-driven templates of inclusive charm decay discriminative variables. In addition, it allows for rejecting a significant fraction of the semileptonic decays from the data without a strong efficiency loss and without strongly biasing muon-only based discriminative variables in simulation. The approach hence preserves to a large extent the discrimination power between prompt and non-prompt dileptons. The latter use of  $\mathtt{SemiCharmTag}$ for background rejection is conceptually close to a boosted decision tree developed for  $B^0_s \to \mu^- \mu^+$ measurements in LHCb~\cite{LHCb:2017rmj}. 
The method takes advantage of the \lhcb detector's capabilities, but the concepts are transferable to any detector with sufficiently good vertex performance, lepton identification and hadron reconstruction. 

In data, another source of background has to be taken into account: the misidentification of hadrons as leptons and combinations of leptons from different sources via multi-parton interactions. The background involving misidentification can be addressed effectively based on dileptons with the same electric charge in data. A possible strategy was employed in the search for dark photons in LHCb~\cite{Dark_photons_LHCb}.  




\section{Simulation data and dimuon selection}
\label{sec:selection}

For this study, $pp$ collisions are generated with the \pythia 8.1 generator~\cite{Sjostrand:2007gs} using a specific \lhcb configuration similar to the one described in Ref.~\cite{Belyaev:1322400}. Decays of particles are described by \evtgen~\cite{Lange:2001uf} in which final-state radiation is generated with \photos~\cite{Golonka:2005pn}. The implementation of the interaction of the generated particles with the detector, and its response, uses the \geant toolkit~\cite{Agostinelli:2002hh,Allison:2006ve}.
The steering of the different steps in the simulation of an event uses interfaces to external generators and to the \geant toolkit. It is handled by \gauss ~\cite{LHCb-PROC-2011-006}, the \lhcb simulation software built on top of the \gaudi~\cite{BARRAND200145,MarcoClemencic_2010} framework. 

The simulation samples correspond to the detector conditions during Run3 2024 $pp$ data taking period at \sqs$=$~13.6~\tev. Three different types of data samples are generated: 
Drell-Yan dimuons ($Z/\gamma^{*}\to\mu^{+}\mu^{-}$) with a mass requirement $M>2$~\gevcc are taken as signal; an inclusive dimuon in the final state from semileptonic decays of charm and of beauty with dimuon \pt~$> 0.9$~\gevc and \ptot $> 9$~\gevc are taken as background sample. 
The candidates are selected according to the requirements listed in Tab.~\ref{tab:hlt2_DY_selections}. 
\begin{table}[hbt]
\centering
\begin{tabular}{l|c|c}
 & Variable               & Requirement\\ 
 \hline
\multicolumn{1}{c|}{Muon}  & \ptot             & $> 10$ \gevc \\
                                & \pt               & $> 1$ \gevc  \\
                                & $\eta$          & $2 < \eta < 5$ \\
                                & $\rm{PID}_\mu$    & $> 0$ \\
                                & IP                & $< 0.4$ \mm \\ 
                                \hline
                                
{Dimuon}                    & $M$              & $2.9 < M < 5$ \gevcc\\
                                & DOCA              & $< 0.15$ \mm  \\
                                & \chisqvtx/DOF     & $< 16$  
\end{tabular}
\caption{List of selections applied to muons from $c\bar{c}$, $b\bar{b}$ and Drell-Yan processes.}
\label{tab:hlt2_DY_selections}
\end{table}
The variables $p, \pt, \eta$ are respectively the momentum, transverse momentum and pseudorapidity of the muon; and $\mathrm{PID}_\mu$ corresponds to the muon identification variable associated with the track.
The DOCA corresponds to the distance of closest approach between the two muons, while \chisqvtx/DOF denotes the vertex quality associated to the dimuon candidate. The fiducial acceptance is defined as the $p, p_{\mathrm{T}}\;\text{and}\; \eta$ selections in Tab.~\ref{tab:hlt2_DY_selections}.

The effect of the IP, DOCA and \chisqvtx/DOF conditions on muon or dimuon levels has been quantified using the simulation generator level information. The requirements reduce the background levels arising from charm (beauty) semileptonic decays by a rejection factor of $34\%$ ($52\%$), while being almost $100\%$ efficient on Drell-Yan events.

\section{Signal and background cross-section estimations}
\label{sec:soverbest}
In order to evaluate the impact of the tagging algorithm in a realistic situation, the initial signal over background ratio, $\mathcal{S}$, has to be computed. We consider background dimuon pairs coming from semileptonic charm and beauty decays and signal dimuons from Drell-Yan as detailed in Eq.~\eqref{eq:signal_background}:

\begin{align}
    \mathcal{S}^c = & \frac{\text{\# signal candidates}}{\text{\# charm background candidates}} = \frac{ \sigma_{p p \rightarrow Z^{0}/\gamma \rightarrow \mu^{+} \mu^{-}} }{\sigma_{pp \rightarrow c \bar{c} \rightarrow \mu^{+} \mu^{-}} },\\
     \mathcal{S}^b = &\frac{\text{\# signal candidates}}{\text{\# beauty background candidates}} = \frac{ \sigma_{p p \rightarrow Z^{0}/\gamma \rightarrow \mu^{+} \mu^{-}} }{\sigma_{pp \rightarrow b \bar{b} \rightarrow \mu^{+} \mu^{-}} },
\label{eq:signal_background}
\end{align}

where $\sigma_{pp \rightarrow c \bar{c}/b \bar{b} \rightarrow \mu^{+} \mu^{-}}$ and $\sigma_{p p \rightarrow Z^{0}/\gamma \rightarrow \mu^{+} \mu^{-}}$ are the dimuon cross-section from charm/beauty and Drell-Yan after the $p, p_{T}\;\text{and}\; \eta$ selection requirements detailed in Tab.~\ref{tab:hlt2_DY_selections}.

The three cross sections in Eq.~\eqref{eq:signal_background} are estimated at Next to Leading Order (NLO) using MadGraph$5\_$aMC$@$NLO~\cite{Alwall:2014hca,Frederix:2018nkq} for the parton level process using the 3 flavour scheme parton distribution function NNPDF30$\_$nlo$\_$as$\_$0118~\cite{NNNPDF30} and \pythia 8~\cite{Sjostrand:2014zea} for the showering, as detailed in Eq.~\eqref{eq:charm_cross_section} and~\eqref{eq:DY_cross_section}.

\begin{equation}
    \sigma_{pp \rightarrow c \bar{c} \rightarrow \mu^{+} \mu^{-}} = \sigma(pp \rightarrow c \bar{c}) \cdot \varepsilon^{c}_{\mathrm{acc}}
\label{eq:charm_cross_section}
\end{equation}

\begin{equation}
    \sigma_{pp \rightarrow b \bar{b} \rightarrow \mu^{+} \mu^{-}} = \sigma(pp \rightarrow b \bar{b}) \cdot \varepsilon^{b}_{\mathrm{acc}}
\label{eq:beauty_cross_section}
\end{equation}

\begin{equation}
    \sigma_{p p \rightarrow Z^{0}/\gamma \rightarrow \mu^{+} \mu^{-}} = \sigma(pp \rightarrow Z^{0}/\gamma) \cdot \varepsilon^{\mathrm{DY}}_{\mathrm{acc}}
\label{eq:DY_cross_section}
\end{equation}

The acceptance factor, $\varepsilon_{\mathrm{acc}}$, in Eq.~\eqref{eq:charm_cross_section}, Eq.~\eqref{eq:beauty_cross_section}, Eq.~\eqref{eq:DY_cross_section}, is defined as the ratio between the dimuons in the fiducial acceptance and the total number of generated events for each process,

\begin{equation}
   \varepsilon_{\rm{acc}} =  \frac{N_{\rm{acc}}({\mu^{+} \mu^{-})}}{N_{\rm{gen}}}.
\end{equation}

The Drell-Yan acceptance factor, $\varepsilon^{\rm{DY}}_{\rm{acc}}$, takes into account the acceptance and kinematic selections on muon and dimuon levels. For the charm and beauty dileptons, this factor also accounts for hadronisation, branching fractions and decay kinematics. The charm cross section is evaluated by starting from dimuons originating from $c\bar{c}$ production hadronizing to $D^0$-$\bar{D^0}$ pairs. The beauty cross-section is based on dimuons from semileptonic decays from $B^+$-$B^-$ pairs, taking into account also intermediate charm mesons. The number of dimuon pairs is then corrected to the number of dimuons of any charm (beauty) hadron anticharm hadron pair, taking into account the combined average fragmentation and decay fractions from LEP $\Gamma (c \to l)/\Gamma (c \to \text{anything})=0.096\pm0.004$ ($\Gamma(b \to l)/\Gamma(b \to \text{anything}) = 0.1069 \pm 0.0022$)~\cite{PDG2024}. For the beauty cross-section, B mixing is not taken into account. The procedure was validated by comparison with single charm fixed-order next-to-leading-logarithm (FONLL) calculations~\cite{Cacciari:2012ny,Cacciari:2015fta}, giving compatible results at the inclusive partonic cross section level. The fiducial cross-section values for each process are: $\sigma_{p p \rightarrow Z^{0}/\gamma \rightarrow \mu^{+} \mu^{-}}  = 6.02\cdot10^{3} \mathrm{pb}$, $\sigma_{pp \rightarrow c \bar{c} \rightarrow \mu^{+} \mu^{-}}  = 8.81\cdot10^{4}\; \mathrm{pb}$ and $\sigma_{pp \rightarrow b \bar{b} \rightarrow \mu^{+} \mu^{-}}  = 9.98\cdot10^{4}\; \mathrm{pb}$. Finally, the resulting signal/background ratio for charm is $\mathcal{S}^{c} = 6.8^{+10.2}_{-4.6}\%$ while for beauty is $\mathcal{S}^{b} = 6.03^{+2.4}_{-2.2}\%$. The uncertainty is given by the charm and beauty total NLO cross section at the parton level. The dominating uncertainties originate from the variation of fragmentation and renormalization scales. Finally, the signal over background ratios within the $p, p_{T}\;\text{and}\; \eta$ selection requirements from Tab.\ref{tab:hlt2_DY_selections} are:
\begin{align}
\label{eq:ratios_prod}
\mathcal{S}^{c} &= 9.1^{+13.7}_{-6.1}\%~ ( {\rm charm})\\ 
\mathcal{S}^{b} &= 9.1^{+3.6}_{-3.3}\%~ ({\rm beauty}).
\end{align}


\section{SemiCharmTag algorithm}
\label{sec:alg}
The $\mathtt{SemiCharmTag}$ algorithm discriminates a prompt lepton from a Drell-Yan candidate (signal) and a non-prompt muon from a semileptonic charm or beauty decay (background). For this purpose, the algorithm forms muon-hadron pairs based on the lepton in question and hadrons reconstructed in the same event. These hadrons and hadron-muon pairs exhibit a different behaviour in topological variables for signal and background candidates, which can be used for discrimination. The algorithm is trained as a three-category classifier with two backgrounds: i.e. charm and beauty. These variables are combined into one-dimensional classifiers that are applied to each hadron-muon pair. Each pair has three classifiers, called probabilities in the following, that are defined: signal probability, charm probability and a beauty probability. A threshold value can be defined in each classifier, and the hadron-muon pair with the lowest probability to be a signal event per lepton is then used to decide whether the lepton is considered background or signal. This logic of classification per lepton can then be used either only on one muon (single-tag) or on both leptons of the dilepton candidate (double-tag). The double-tag is used primarily to improve the signal over background, whereas the single-tag is used to construct pure samples of unbiased background leptons, selecting only the leptons with a high probability of coming from charm.


\subsection{Discriminant variables}\label{Discriminant_variables_section}
The separation between muons from semileptonic decays of charm or beauty, and prompt muons is based on properties of the secondary vertex as well as the hadron track associated with the muon. The hadron track from a charm decay, a beauty decay and a hadron track from the underlying event of a Drell-Yan or thermal/preequilibrium dilepton muon differ. The hadron track from charm is non-prompt, it stems from the same decay vertex as the muon and it is more likely to be a kaon. In addition, it is likely to be of opposite charge than the muon, is harder in transverse momentum than the underlying event and forms an invariant mass with the muon that is smaller than the charm hadron mass. The hadron track from beauty is typically better separated from the primary vertex than the charm decay track and exhibits a larger invariant mass than the charm decay track.   
Therefore, variables related to distance measurements between the reconstructed tracks and vertices,  probabilities that the muon and associated hadron come from the same vertex, invariant mass associated with the combination, particle identification information as well as kinematics of the tracks are particularly useful for background rejection. 

In the following, a detailed list of variables used for discrimination is presented:
\begin{itemize}
    \item \textbf{Impact parameter (IP)}: The impact parameter is defined as the distance of closest approach between the reconstructed track and the closest PV.
    \item \textbf{Impact parameter $\boldsymbol{\chisq}$ ($\boldsymbol{\chisqip}$)}: The \chisqip of a particle track with respect to a given vertex (usually the PV) is defined as the increase in the vertex-fit \chisq when the track is added to the vertex. It quantifies how compatible the track is with originating from that vertex.
    \item \textbf{Distance of closest approach (DOCA)}: Defined as the smallest distance between two decay products, in our case between the muon and the hadron track at the decay vertex.
    \item \textbf{Direction of flight angle (DIRA)}: Angle between a line drawn from the closest PV to the decay vertex of the particle and the sum of the 4-momentum of its decay products. 
    \item \textbf{Flight distance (FD)}: Distance between the interaction point and the decay vertex of the particle.
    \item \textbf{Flight distance} \textbf{$\boldsymbol{\chisq}$ ($\boldsymbol{\chisqfd}$)}: \chisq of the flight distance of a particle with respect to a vertex. A measure of how well the decay vertex of the particle can be separated from a vertex. 
    \item \textbf{Particle identification Kaon ($\boldsymbol{\rm{PID_K}}$)}: Probability of the associated track to be a kaon using particle identification (PID) variables, which uses the differences in the likelihoods between kaon and pion mass hypotheses for calculation.
    \item \textbf{Transverse momentum ($\boldsymbol{\pt}$)}: Transverse momentum of the associated track. 
    \item \textbf{Invariant mass ($\boldsymbol{M}$)}: Invariant mass of combined muon and hadron.
\end{itemize}

For charm tagging, variables directly related to the muons or muon pairs are not used in order to limit bias for the signal extraction with template fitting.  


\subsection{Implementation}

The tagging technique is implemented using the \davinci software \cite{web:davinci} in \lhcb, adding a tagging charged particle to both muons in the dimuon pair via a hypothetical resonance $D^{0}\to\mu^{+}\pi^{-}$ and charged conjugate. Preselections, based on DOCA($\mu$,\hadron), $M$($\mu$, \hadron) and $\pt^{\hadron}$, are applied on the tagging hadron track and muon-hadron pair to reduce the number of combinations on the single muon level.  The selections are reported in Tab.~\ref{tab:preselections_D0_table}.

\begin{table}[htpb]
\centering
\begin{tabular}{c|c}
Variable & Requirement\\ 
\hline
DOCA($\mu$, \hadron)      & $< 0.4 $ \mm                 \\
$M$($\mu$, \hadron)      & $140 < M(\mu, \hadron) < 2900 $ \mevcc      \\
$\pt^{\hadron}$           & $> 500$ \mevc
\end{tabular}
\caption{Selection requirements applied, where \hadron is the hadron and ($\mu$, \hadron) the muon-hadron track pair.}
\label{tab:preselections_D0_table}
\end{table}
Additional requirements are applied for muon identification~\cite{Anderlini:2020ucv} and ghost rejection at the track level.

In the case of the Drell-Yan signal, the tagging hadron tracks are mostly prompt pions from the primary vertex associated with the prompt muons.
The muon-hadron vertex is a combinatorial construction of two prompt tracks. In the case of the charm (beauty) background,  one or more of the hadron-muon pairs correspond to the real decay vertex of the charm (beauty) hadron. 

Machine-learning approaches are essential to take full advantage of the rejection power of the previously described quantities among the different charm species, as each decay chain induces different correlations between the observables.  
The variables detailed in Sec.~\ref{Discriminant_variables_section} are used as input for training a Boosted Decision Tree (BDT) using the XGBoost~\cite{Chen:2016btl} framework.

After applying the selection requirements described previously, a generator-level selection is performed before using the signal and background samples for BDT training, with the goal of rejecting hadrons misidentified as muons. 
For the signal (Drell-Yan), muons are selected originating from a $Z/\gamma^{*}\to\mu^{+}\mu^{-}$ decay. For the background, semileptonic charm and beauty decays are selected with generator-level information. For the charm sample, at least a muon and one additional track are required to be reconstructed and originating from a common secondary charm vertex.
The semileptonic beauty decays are treated analogously to the charm case. We define, in this context, background from beauty decays as the background of muons arising directly from the weak decays of beauty hadrons, or indirectly from the decay of intermediate charm hadrons.

Regarding the background simulations, only the charm species listed in Tab.~\ref{tab:charm_parents_table} are taken into account for charm hadrons produced at the PV and charm from beauty decays. The simulation requires that the $c\bar{c}$ and $b\bar{b}$ pairs decay into a muon in the LHCb acceptance.



\begin{table}[htb]
\centering

\begin{tabular}{|c|c|c|}
\hline
 & $\# \mu^+$ & $\# \mu^-$   \\
\hline
\Dz      & $53$k & $53$k  \\
\Dp      & $76$k & $76$k   \\
\Dsp     & $20$k & $20$k   \\
\Lc      & $7$k  & $7$k  \\
\Xicp    & $37$  & $26$ \\
\Xic     & $316$ & $283$ \\
\Omegac  & $7$   & $6$ \\
\hline
\end{tabular}
\caption{Semileptonically decaying charm hadron species from the inclusive charm simulation sample. The total number of $\mu^+$ and $\mu^-$ is provided without selections requiring that the muon has an associated hadron track.    
}
\label{tab:charm_parents_table}
\end{table}

\begin{figure}[htb]
    \centering
    \includegraphics[width=0.95\linewidth]{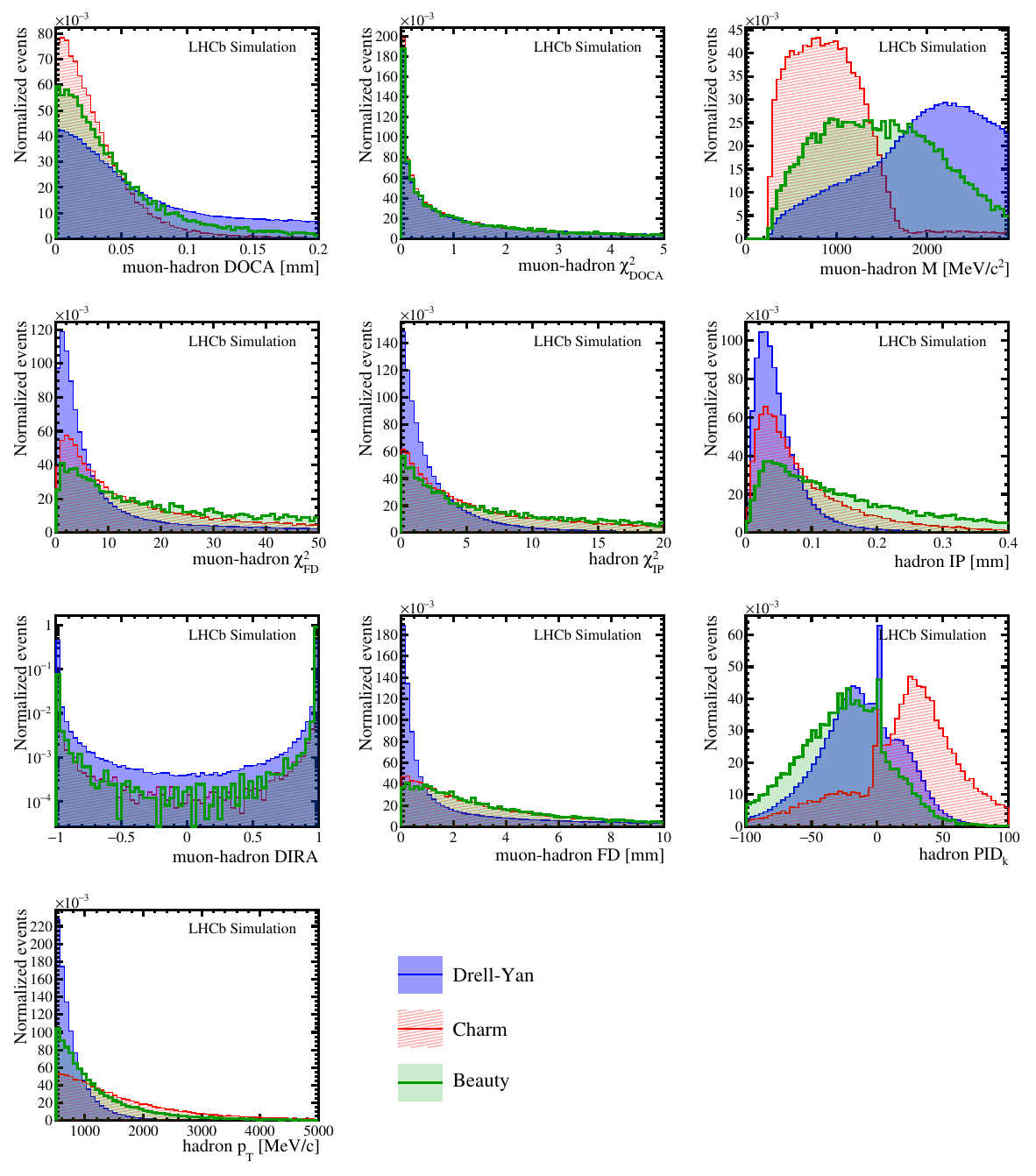}
    \caption{Distributions of the different discriminant variables listed in Sec.~\ref{Discriminant_variables_section}, normalised to unity, for Drell-Yan (blue), charm (red) and beauty (green) simulation samples. 
    }
\label{fig:discriminant_variables_fig}
\end{figure}

Signal and background samples are divided into independent training and testing subsamples with a 70/30 training-to-testing ratio. The full charm background sample contains 57000 candidates, the full beauty background sample contains 25000 candidates and the signal sample 550000 candidates. To compensate for this sample-size imbalance and prevent the XGBoost classifier from being dominated by the signal sample during training, a class weight $w_c = N_{\rm{total}}/({N_c \cdot N_{\rm{classes}}})$ is applied during training, with ${N_{\rm{total}}}$ the total number of training events, $N_c$ the number of events in a given class and $N_{\rm{classes}}$ the number of classes. This class weight allows to equalise the contributions of each class to the loss function. The BDT is trained using the input variables on the preselected simulation samples with 300 estimators, a learning rate of 0.1 and a maximum tree depth of 5. The distributions of the discriminative variables used for training are shown in Fig.~\ref{fig:discriminant_variables_fig} for the three sample categories. 
\clearpage
\subsection{SemiCharmTag performance on the hadron-muon pair}
\begin{figure}[htb]
    \centering
    {
   \includegraphics[width=0.5\linewidth]{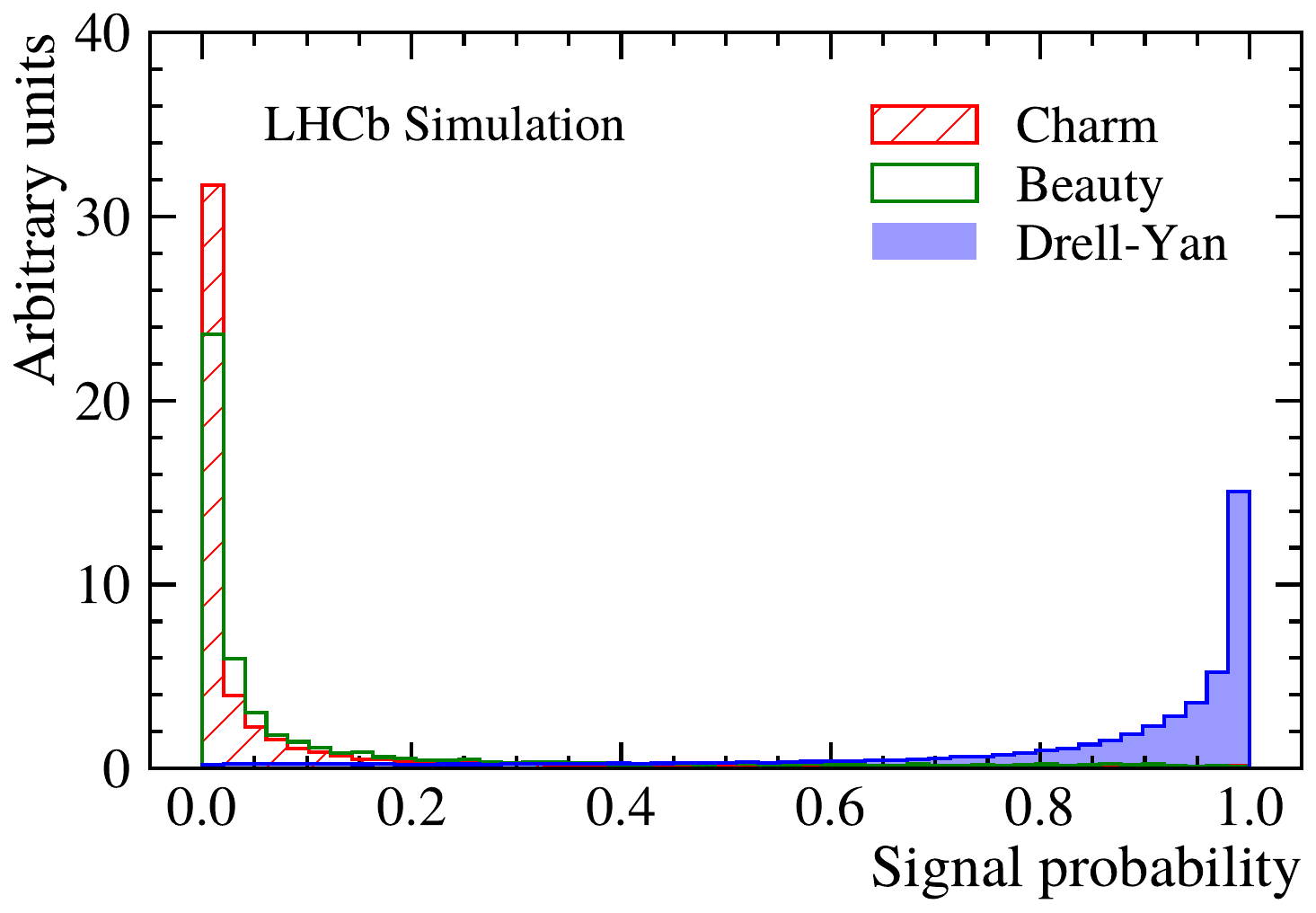}
    }
    {\includegraphics[width=0.43\linewidth]{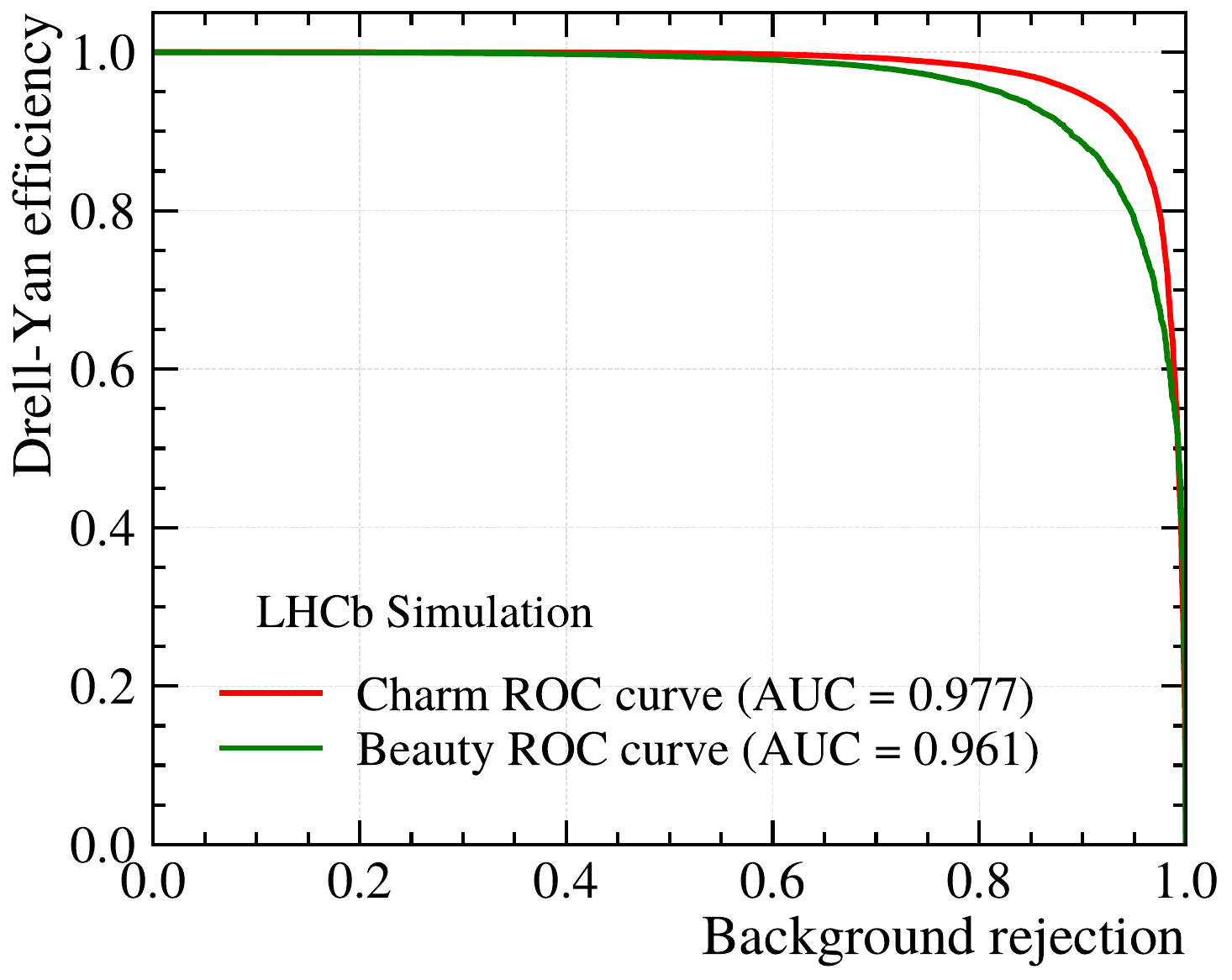}
    }
    \caption{Left: Signal probability score for signal (Drell-Yan, blue) and background (charm, red and beauty, green). Right: Separation of signal and background in the classifier, ROC curve, where each entry is a muon-hadron pair.}
    \label{fig:BDT_performance}
\end{figure}

In the following, the performance is shown in the subsample of the background muons that exhibit a secondary vertex with at least one charged track passing the preselection specified before both for the charm as well as the beauty sample. 
The BDT performance in separating signal and background components is shown in Fig.~\ref{fig:BDT_performance} as a function of the signal probability on the left-hand side. The receiver-operating characteristic (ROC) curve is shown for the charm and the beauty rejection with the corresponding area under the curve (AUC) in Fig.~\ref{fig:BDT_performance} on the right-hand side.  

By choosing a BDT signal probability requirement of 0.15, around 90\% of muons from charm decays and 80\% of muons from beauty decays are rejected with a 96\% signal efficiency. The performances demonstrate that in the presence of a hadron track at the secondary vertex, the separation of signal and background is almost completely achieved based on the BDT.  

\section{Double-tag for background rejection}
\label{sec:double}

In a dimuon event passing the selection criteria detailed in Sec.~\ref{sec:selection}, a requirement based on the BDT output can be applied either on the $\mu^+$ and the $\mu^-$ (double-tag) or only on one of the two muons (single-tag). 

\begin{figure}[tb!]
    \centering
    \includegraphics[width=0.65\linewidth]{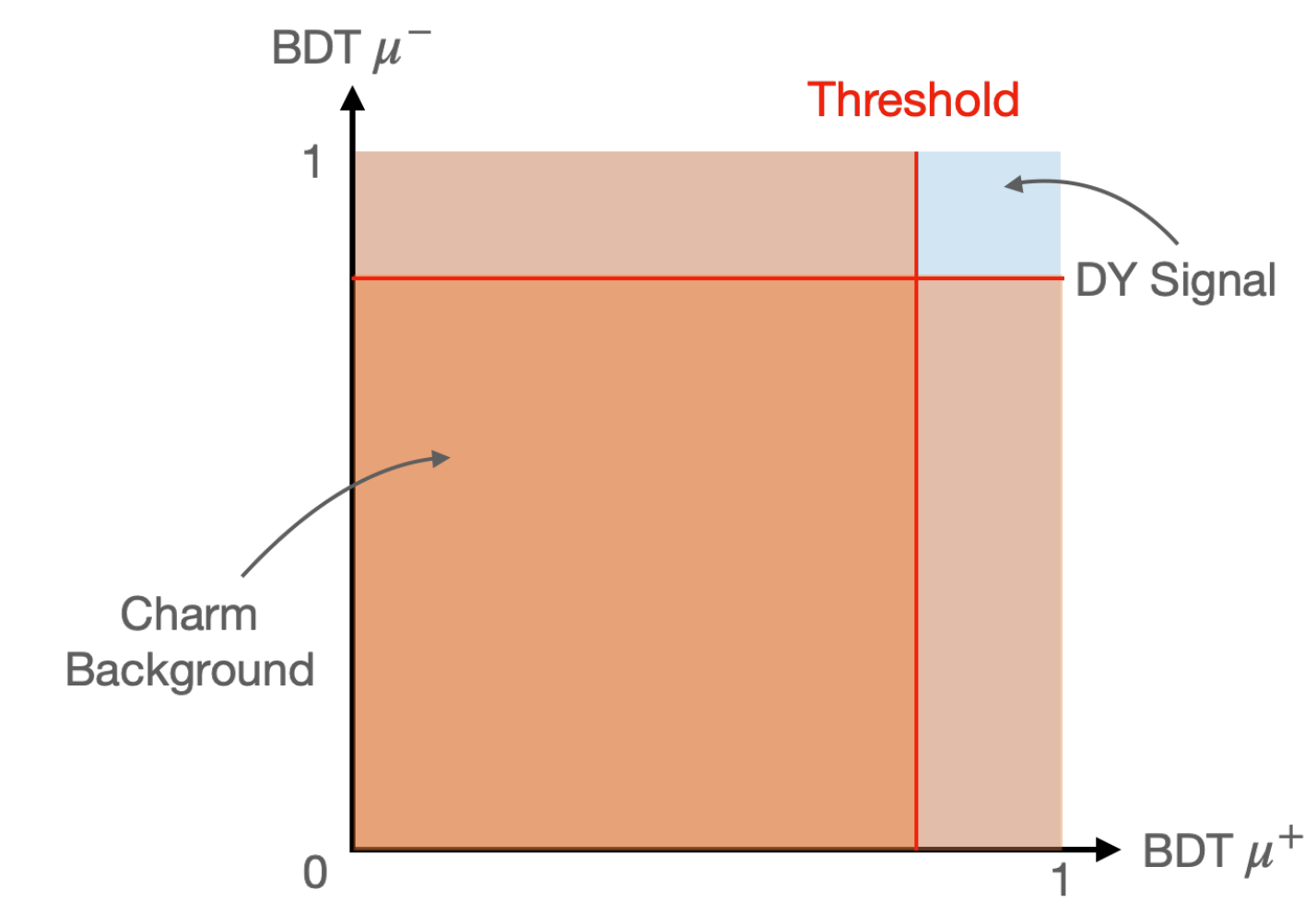}
    \caption{Illustration of the double-tag strategy. The threshold is shown in red. If at least one muon-hadron pair in a dimuon candidate is below a defined threshold, the candidate is classified as charm or beauty and rejected, as shown in the orange region. Hence, both muons should pass the threshold to be classified as signal, as illustrated in light blue.}
    \label{fig:double_tag_cartoon}
\end{figure}

In order to achieve the maximum rejection of the charm and beauty background for prompt dilepton analyses, the double-tag strategy provides the highest background rejection power.
The double-tag requirement is defined as follows: if at least one muon–hadron pair in a dimuon candidate falls below a predefined signal probability threshold, the candidate is classified as originating from charm or beauty and is rejected. Otherwise, if both muons pass the threshold, they are classified as signal. The method is illustrated in Fig.~\ref{fig:double_tag_cartoon}.
In the following, the performance of the double-tag strategy and its limitations are discussed. 

\subsection{Natural limitations}
The double-tag strategy performance is naturally limited by two factors. First, the separation power between hadrons from charm decays and primary hadrons. Second, the fraction of semileptonic charm decays with detectable hadron tracks. Similar considerations apply to beauty decays. We focus here on the charm case since it is more difficult to separate based on secondary vertex information due to the smaller lifetimes. 



A significant fraction of semileptonic decays do exhibit only a lepton, a neutrino and a long-lived neutral particle that are not easily attached to the decay vertex. A prototypical example of this case is the following decay chain $D^+ \to \mu^+ \nu_\mu \bar{K^0}$.  Moreover, the number of charged tracks that are reconstructed in all tracking subdetectors is limited by the detector acceptance, the interaction of the tracks in the material, and the efficiency of the reconstruction algorithm. 
Within the employed simulation, 65\% of all charm background dimuon candidates exhibit at least one charged hadron at one of the two charm decay vertices. 
This percentage could be improved by considering also hadron tracks with the same charge as the muon track and by considering also tracks that are only reconstructed in the vertex detector, enlarging the acceptance and including some tracks lost downstream due to hadronic interactions with the detector material.

The tagging algorithm also has a smaller efficiency for background than for signal in situations where no hadron track is reconstructed at the charm decay vertex. Since the discriminative variables include the transverse momentum of the hadron and the mass between the muon and the hadron, this behaviour can be expected since a parton shower is typically attached to the charm production, whereas the colour-neutral Drell-Yan events exhibit less hadronic activity in the vicinity of the dilepton. In principle, the previous differences could be used by dedicated machine learning methods, adding other variables targeting lepton isolation, but the differences in simulation between signal and background may not fully reflect the situation in data.
The signal efficiency is, however, difficult to control in data, since no pure source of colour-neutral Drell-Yan production can be easily isolated in data for validation. Therefore, we choose in this simulation-based study a signal efficiency operational point of $81\%$ in the following for illustration. 





\subsection{Background rejection performance}

\begin{figure}[htb]
    \centering
    \includegraphics[width=0.5\linewidth]{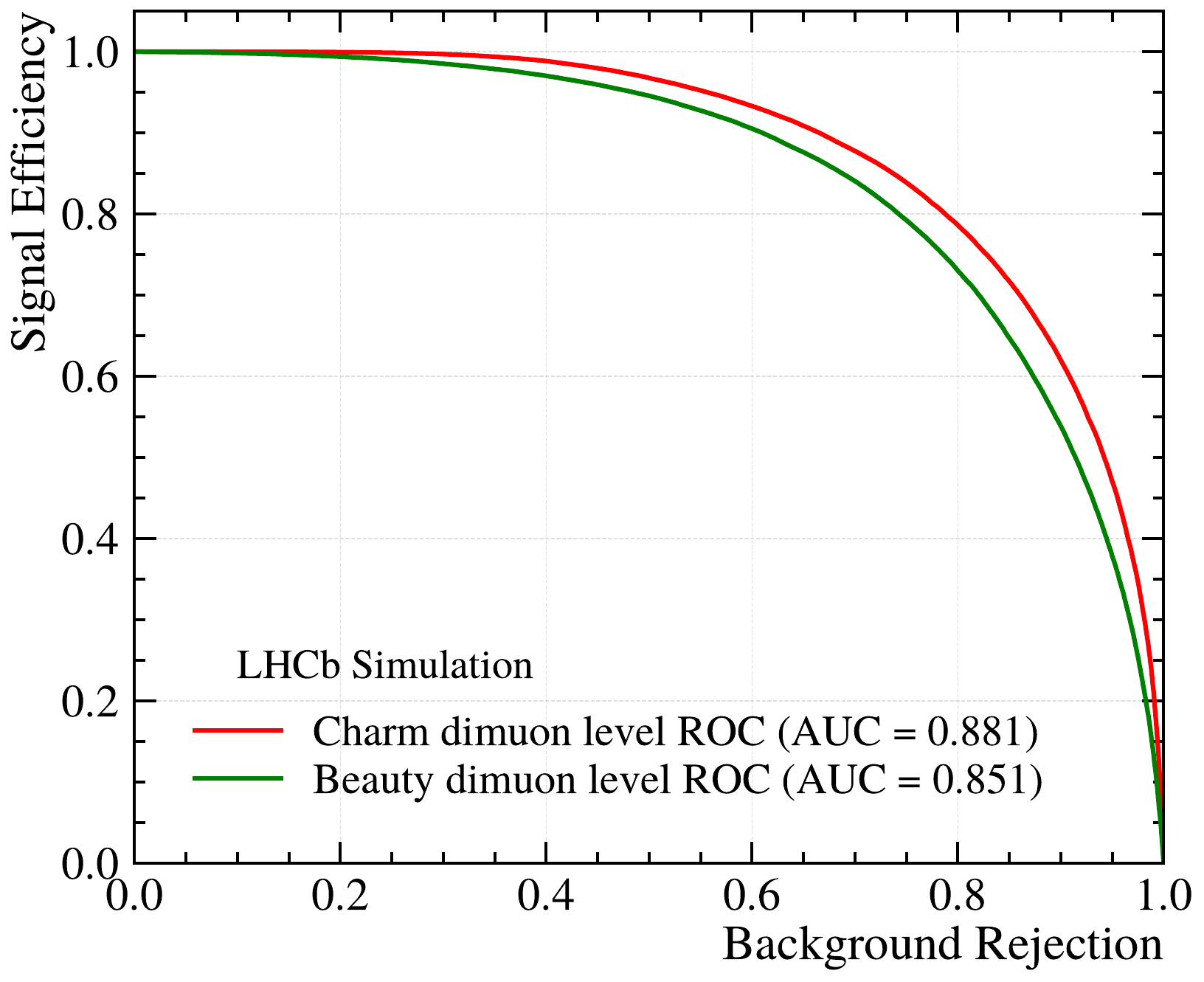}
    \caption{Performance of the tool for signal efficiency as a function of charm (red) and beauty (green) background rejection using the double-tagging strategy on an inclusive charm sample. }
    \label{fig:dimuon_ROC_curve}
\end{figure}

The double-tag strategy, described in the application strategy Sec.~\ref{sec:double}, classifies a dimuon event as background if at least one muon-hadron pair has a BDT signal probability below a chosen threshold. The probability of producing and reconstructing at least one charged hadron track passing the preselection at the charm decay vertex of a muon from a semileptonic decay is around $65 \%$ in simulation dimuon events from charm production. The application of a double-tag strategy where the rejection criteria are used independently on both muons improves the overall background rejection. The signal efficiency as a function of the charm background rejection is detailed in Fig.~\ref{fig:dimuon_ROC_curve}. The ROC curve shows the dimuon-level background rejection using a double-tag strategy. 
At the chosen operational point, this results in a signal efficiency of $81\%$ at a charm (beauty) rejection of 78\% (74\%), improving the signal over background ratio by a factor 3.7 (3.2).



\subsection{Bias on the signal and background samples}

\begin{figure}[tb!]
\centering
  \includegraphics[width=0.48\textwidth]{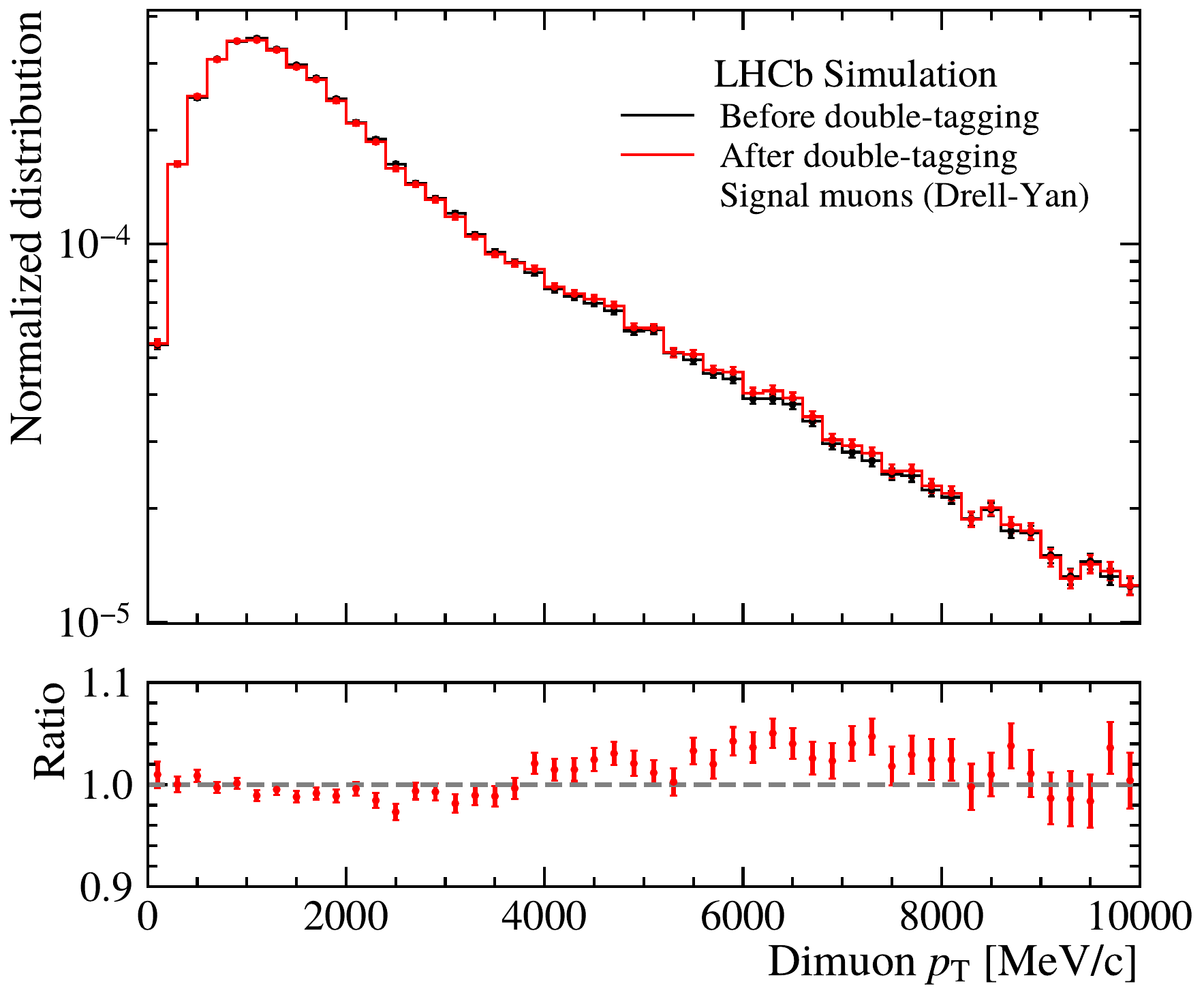}
  \includegraphics[width=0.48\textwidth]{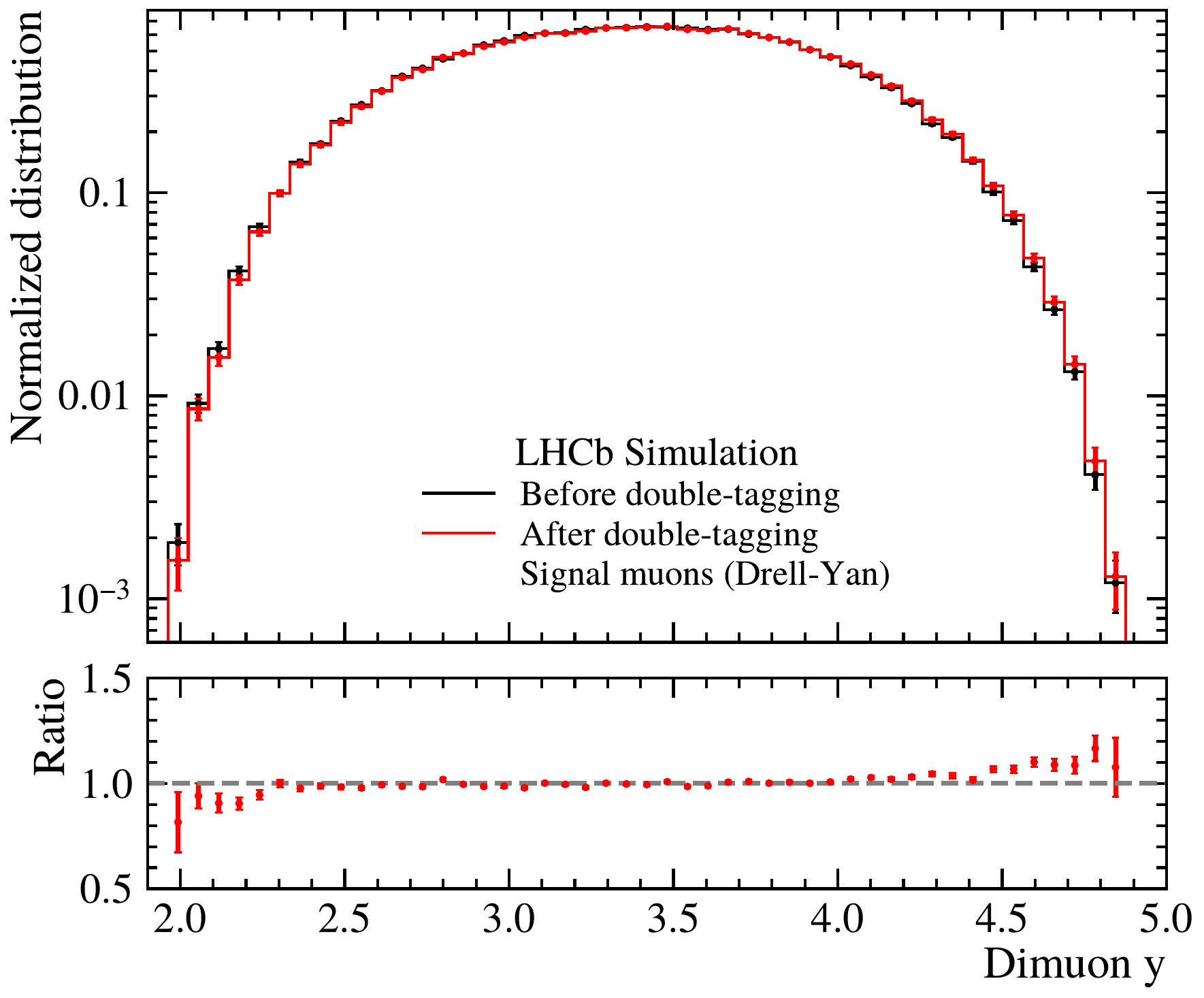}
\caption{Normalized $\pt$ (left panel) and rapidity (right panel) distributions for Drell-Yan dimuon candidates before applying the double-tagging strategy (black) and after selection (red), as well as the ratio between both distributions.}
\label{fig:bias_background_rejection}
\end{figure}

Since the tagging does not allow for the removal of the semileptonic charm background completely, a signal extraction based on other discriminant variables has to be performed after the application of the tagging selection. As explained in the introduction, variables related to the displacement of one or both muons are most suited for this purpose. 
For the extraction of the cross sections, it is important to quantify how much the kinematic variables of interest are biased. 
We evaluate the distribution of the Drell-Yan dimuon \pt and the Drell-Yan dimuon rapidity in simulations before and after the double-tag in Fig.~\ref{fig:bias_background_rejection}.
The deviations in the normalized dimuon \pt distributions before and after the double-tagging are below 5\%. The deviations for the normalized dimuon rapidity distributions are of similar size, with the exception of the acceptance edges. The deviations between y$\in [2,2.3] $ and y$\in [4.4,5]$ can reach up to 25\%.

\begin{figure}[ht!]
\centering
  \includegraphics[width=0.48\textwidth]{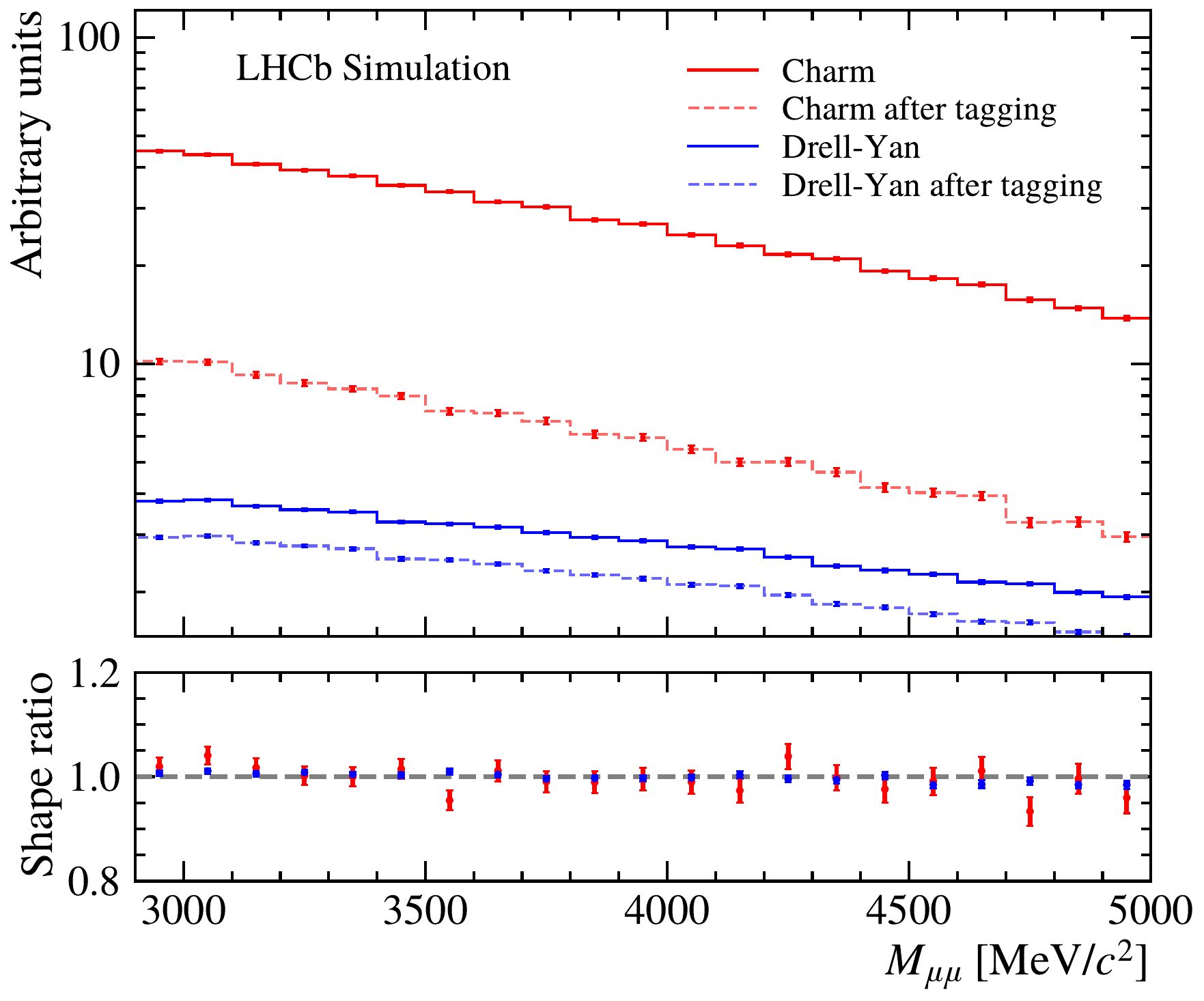}
  \includegraphics[width=0.48\textwidth]{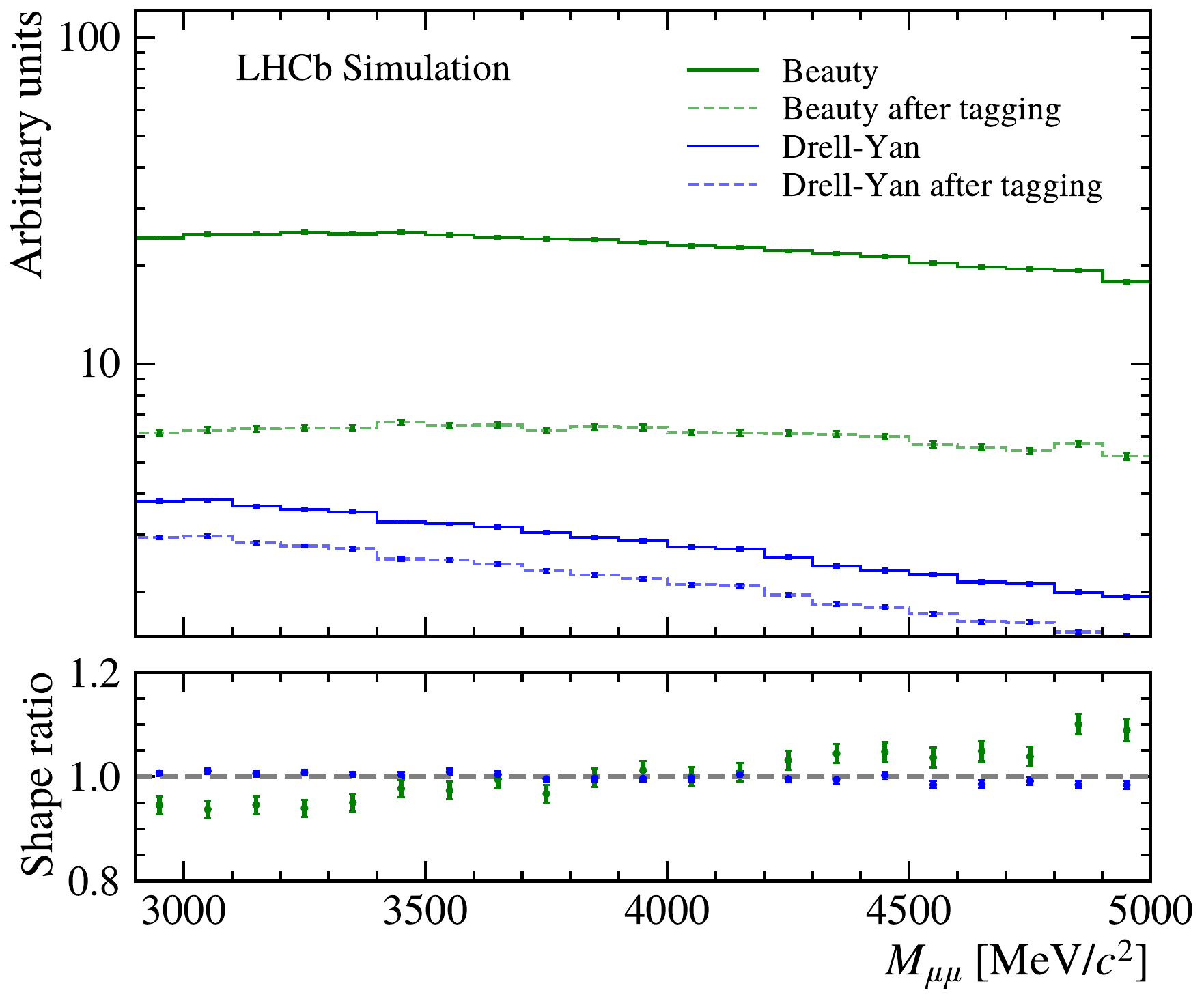}
\caption{Top: Dimuon mass distribution for Drell-Yan (blue), charm (red) and beauty (green) before applying the double-tag strategy (full line) and after selection (dashed line). Bottom: Shape ratio of the mass distribution before and after tagging.}
\label{fig:mass_distribution_after_BDT}
\end{figure}

The effect of the tagging technique on the mass distributions for prompt and non-prompt dimuons is reported in Fig.~\ref{fig:mass_distribution_after_BDT}. 
The Drell-Yan, charm and beauty distributions are both normalized and scaled by their corresponding cross-section computed in Sec.~\ref{sec:soverbest} to show the relative reduction of background and signal after the selections. There is no notable modification in the shape of the signal distributions compared to before the application of the tool. The shape of the charm background does not show any deviations beyond the 5\% level, whereas the beauty background is slightly better rejected at low invariant masses compared to higher masses.

\section{Single-tag for template construction}
\label{sec:single}

Given the limited knowledge on charm hadron production and decays, it is important to establish a method to extract a pure sample of leptons originating from charm decays in data, as explained in Sec.~\ref{sec:Introduction}.

The double-tag strategy cannot be used to retain unbiased template distributions: the kinematic variables of the hadron are correlated with the muon ones, and the fractions of specific hadron species contributing to tagged charm hadron decays compared to the inclusive sample of semileptonic decays are different. Since the charm hadron lifetimes are considerably different, this leads to biases in the variables of displaced charm. Fig.~\ref{fig:bias_background_rejection} demonstrates this behavior.

\begin{figure}[htpb]
    \centering
    \includegraphics[width=0.95\linewidth]{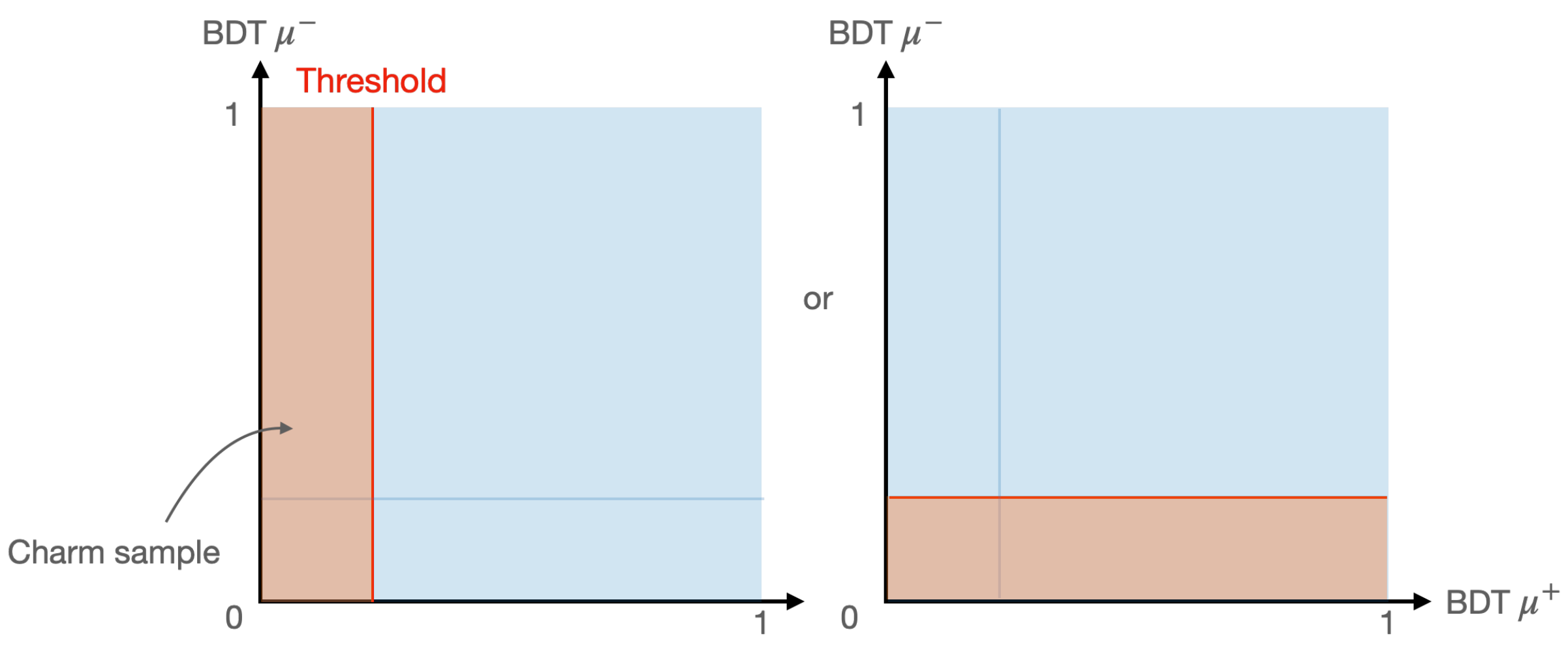}
    \caption{Illustration of the single-tag strategy. A tagging muon below a defined threshold, as illustrated in the orange region, is used to probe the other muon and construct an unbiased pure charm sample.}
    \label{fig:single_tag_cartoon}
\end{figure}

Instead, a single-tag strategy can be applied similar to the strategy used for the calibration of the c-jet tagging in LHCb~\cite{LHCb:2021dlw}. In this case, we require a strict selection on the first muon or antimuon, and we extract the lepton characteristics, such as IP and/or other variables on the second muon. In the case of a 3-category classifier, this can be achieved by requiring a high charm BDT probability, while requiring a low beauty probability to limit the contamination of beauty.
The illustration reported in Fig.~\ref{fig:single_tag_cartoon}  describes the single-tag strategy. The procedure enables the extraction of unbiased single muon distribution observables while suppressing the prompt muons. Based on the single muon information, dimuon information can be constructed based on the momentum and $\eta$ dimuon kinematics in a double-tag sample or in simulation. 

\subsection{Natural limitations}

The strategy considered for the template construction consists of the extraction of single muon properties, the probe muons, with the single-tag algorithm. In order to construct a dimuon template, the single tag algorithm is used to extract $\mu^+$ and $\mu^-$ properties. With these two samples of single muons and with additional information on pair kinematics from simulation or double-tagging, the dimuon template for charm is constructed. 

The performance of the single-tag algorithm to extract templates with this procedure is determined by three factors. The efficiency for semileptonic decays, the rejection of prompt leptons and the bias induced on the probe muon side while selecting with the tagging side. In addition, the contamination from semileptonic decays of beauty is relevant in the kinematic domains considered in this work, detailed in Tab.~\ref{tab:hlt2_DY_selections}. The resulting contaminations and biases on single muon level  enter both for the $\mu^+$ and the $\mu^-$ characteristics independently.  

The fraction of dimuons in the sample from charm, beauty, and Drell-Yan as $\mathcal{C}$, $\mathcal{B}$, and $\mathcal{DY}$ prior to any tagging  is defined as:

\begin{align}
\mathcal{C} &= \frac{N_{\mu\mu}^{c\bar{c}}}{N_{\mu\mu}^{c\bar{c}} + N_{\mu\mu}^{b\bar{b}} + N_{\mu\mu}^{\mathrm{DY}}} , \\ 
\mathcal{B} &= \frac{N_{\mu\mu}^{b\bar{b}}}{N_{\mu\mu}^{c\bar{c}} + N_{\mu\mu}^{b\bar{b}} + N_{\mu\mu}^{\mathrm{DY}}} , \\ 
\mathcal{DY} &= \frac{N_{\mu\mu}^{\mathrm{DY}}}{N_{\mu\mu}^{c\bar{c}} + N_{\mu\mu}^{b\bar{b}} + N_{\mu\mu}^{\mathrm{DY}} },
\end{align}
where $N_{\mu\mu}^{c\bar{c}}$, $N_{\mu\mu}^{b\bar{b}}$, $N_{\mu\mu}^{\mathrm{DY}}$ are dimuon pairs from charm-charm, beauty-beauty and Drell-Yan.  

With these expressions, we define the single tag efficiencies as:
\begin{align}
\text{Charm}:&~\epsilon_{\rm{charm},{\mu^\pm}}= \frac{N_{\mu\mu}^{\mathrm{c\bar{c}}}|_{\rm post~selection,~tag~on~\mu^{\mp}}}{N_{\mu\mu}^{\mathrm{c\bar{c}}}},\\
\text{Beauty}:&~ \epsilon_{\rm beauty,\mu^\pm}= \frac{N_{\mu\mu}^{\mathrm{b\bar{b}}}|_{\rm post~selection,~tag~on~\mu^\mp}}{N_{\mu\mu}^{\mathrm{b\bar{b}}}},\\
\text{Drell-Yan}:&~\epsilon_{\rm{DY,\mu^\pm}}=\frac{N_{\mu\mu}^{\mathrm{DY}}|_{\rm post~selection,~tag~on~\mu^\mp}}{N_{\mu\mu}^{\mathrm{DY}}}. 
\end{align}

  Based on these efficiencies, we define the total efficiency for the single-tag algorithm as:
\begin{align}
 &   \epsilon_{\mathrm{tot},~\mu^{\pm}} = \mathcal{C}\cdot\epsilon_{\rm{charm,~\mu^\pm}} +  \mathcal{B}\cdot\epsilon_{\rm{beauty,~\mu^\pm}} + \mathcal{DY}\cdot\epsilon_{DY,~\mu^\pm}
   \end{align}
   We assume that the efficiencies and contaminations for the probe muon are independent of the muon charge:
   \begin{align}
& \epsilon_{\mathrm{charm}} \simeq \epsilon_{\mathrm{charm},~\mu^+} \simeq \epsilon_{\mathrm{charm},~\mu^-},\\
& \epsilon_{\mathrm{beauty}} \simeq \epsilon_{\mathrm{beauty},~\mu^+} \simeq \epsilon_{\mathrm{beauty},~\mu^-},\\
& \epsilon_{\mathrm{DY}} \simeq \epsilon_{\mathrm{DY},~\mu^+} \simeq \epsilon_{\mathrm{DY},~\mu^-},\\
& \epsilon_{\mathrm{tot}} \simeq \epsilon_{\mathrm{tot},~\mu^+} \simeq \epsilon_{\mathrm{tot},~\mu^-}.
\end{align}


Once the $\mu^+$ and $\mu^-$ properties are extracted separately using the single-tag algorithm, the two single-muon samples are combined into a dimuon template. The dimuon information is constructed using momentum and $\eta$ kinematics derived from either a double-tag sample or simulation. The dimuon template contains a specific fraction of true charm dimuon pairs:
\begin{align}
\label{eq:TrueCharmFract}
    C\bar{C}& = \mathcal{C}^2 \cdot \epsilon_{charm,\mu^+}/\epsilon_{tot,\mu^+} \cdot \epsilon_{charm,\mu^-}/\epsilon_{tot,\mu^-} \simeq \mathcal{C}^2 \cdot (\epsilon_{charm}/\epsilon_{tot})^2.
\end{align}
The dimuon template extracted from the single-tag contains hence prompt-prompt, prompt-charm,  beauty-beauty, prompt-beauty, charm-beauty contaminations, $C_{i}$ listed below:

\begin{align}
\label{eq:cont}
 \text{prompt-prompt}:   C_{pp} & \simeq \mathcal{DY}^2 \cdot (\epsilon_{\rm{DY}}/\epsilon_{\mathrm{tot}})^2   \\
  \text{prompt-charm}:   C_{p\bar{c}} & \simeq C_{cp} \simeq\mathcal{DY} \cdot \epsilon_{\rm{DY}}/\epsilon_{\mathrm{tot}} \cdot \mathcal{C} \cdot \epsilon_{\rm{charm}}/\epsilon_{\mathrm{tot}}   \\
  \text{beauty-beauty}:   C_{b\bar{b}} & \simeq \mathcal{B}^{2} \cdot (\epsilon_{\rm{beauty}}/\epsilon_{\mathrm{tot}})^{2} \\
 \text{charm-beauty}:   C_{c\bar{b}} & \simeq C_{b\bar{c}} \simeq \mathcal{C} \cdot \epsilon_{\rm{charm}}/\epsilon_{\mathrm{tot}} \cdot\mathcal{B} \cdot \epsilon_{\rm{beauty}}/\epsilon_{\rm{tot}} \\
 \text{prompt-beauty}:   C_{p\bar{b}} & \simeq C_{bp} \simeq\mathcal{DY} \cdot \epsilon_{\mathrm{DY}}/\epsilon_{\mathrm{tot}} \cdot \mathcal{B} \cdot \epsilon_{\mathrm{beauty}}/\epsilon_{\mathrm{tot}},
\end{align}
where $C\bar{C}+\sum_i C_i = 1 $. \\

The size of the contaminations and their uncertainties represent a source of systematic uncertainty. 

\subsection{Template construction performance}

The ideal selection is a trade-off between residual contamination from Drell-Yan and beauty muons inside the pure probe charm sample, small selection biases and the limitation of statistics loss. In particular, we target a $C_{pp}$ contamination well below $1\%$ and a statistical reduction by less than a factor 10 to reduce the direct bias on the extraction of prompt dileptons in the signal and minimize the statistical uncertainty of the charm sample. Following these criteria,  a selection resulting in a charm efficiency of $21.4\%$ at a Drell-Yan efficiency of $1.1\%$ and a beauty efficiency of $2.3\%$ on the single muon level is chosen. 

\begin{figure}[htb]
    \centering
    \includegraphics[width=0.6\linewidth]{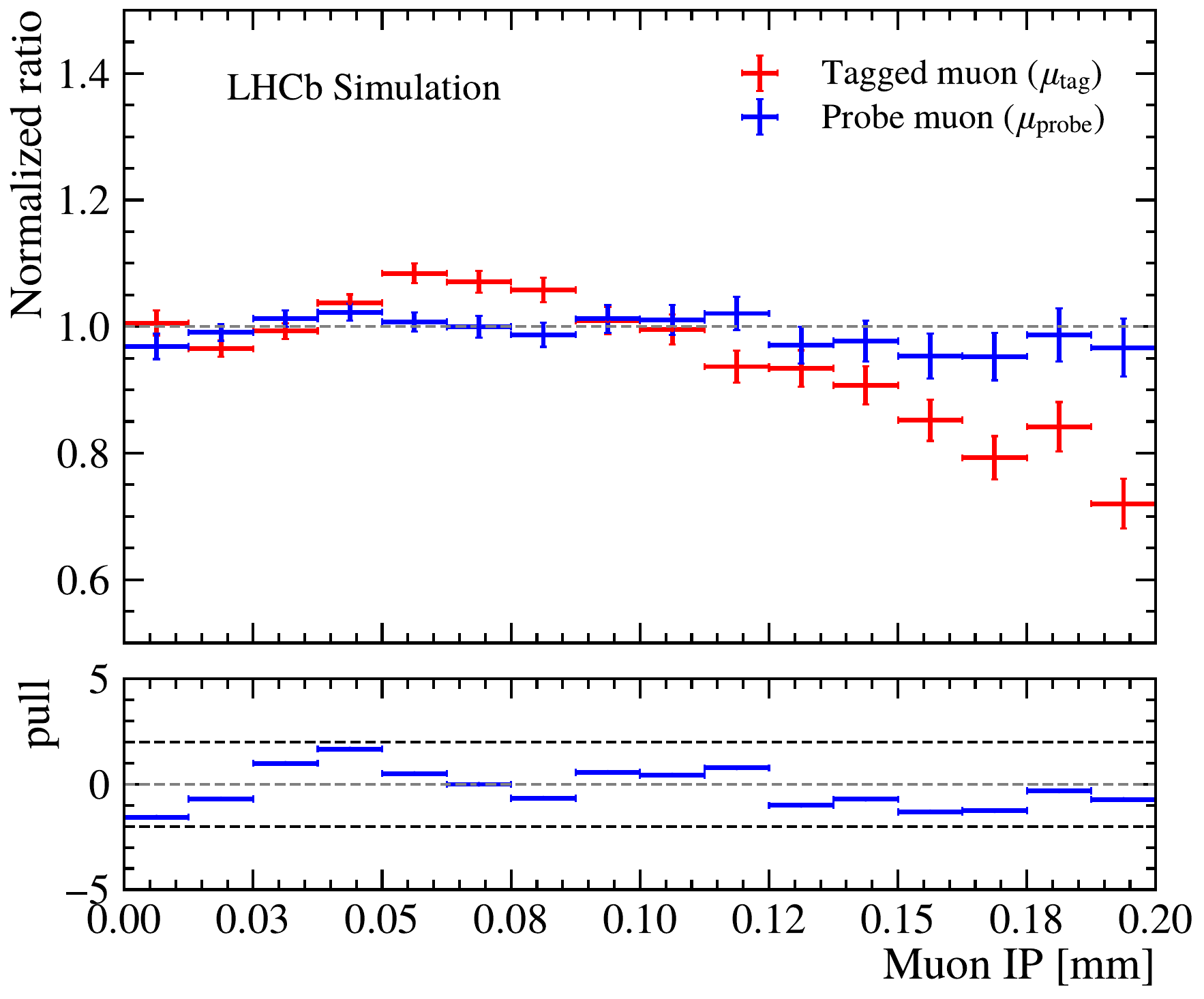}
    \caption{Ratio of normalized IP distributions before and after applying the single-tag strategy, shown for the tagged muon (red) and probe muon (blue) at the chosen BDT operational point. The bottom panel shows the pulls, defined as $(R-1)/\sigma_R$, with $R$ the ratio and $\sigma_R$ its statistical uncertainty.}
    \label{fig:tagprobe}
\end{figure}

For this operational point, we verify in simulations first that the single-tag method is not producing a biased template in the absence of any contaminations.  For this purpose, the ratio of the IP distribution after applying the single-tag strategy compared with the full single-lepton sample is shown in Fig.~\ref{fig:tagprobe}. As can be observed, the ratio is consistent with unity within $2\sigma$, showing no strong bias on the probe muon. In addition, no bias on the transverse momentum and the pseudorapidity distribution of the probe was observed in the simulation. The observed bias can be understood by the observed correlation between hadron IP and related observables in particular important for the separation between the charm and beauty samples with respect to the muon IP on the tag side that is correlated with the muon IP on the probe side.  

Based on the cross-section ratios detailed in Sec.~\ref{sec:soverbest} and the expressions from Eq.\eqref{eq:cont} and following, we can estimate the approximate size of the contamination in a dimuon template constructed from the single-muon distributions of the probe muons to be expected in data. We consider two case studies. In the first, a charm-to-beauty production ratio of 0.5 is assumed, while in the second, a ratio of 2 is taken into account. These values bracket the production ratio in Eq.~\eqref{eq:ratios_prod}.
The second choice is expected to be closer to the situation in real data since the production cross sections for charm and beauty at the LHC are typically at the upper bound of next-to-leading order calculations, see e.g.~\cite{LHCb:2015swx,LHCb:2017vec}. This suggests a larger charm over beauty production cross section than assumed here so far, given the larger uncertainties on the charm production. 
The most critical contribution is the residual contamination of prompt dimuons ($C_{pp}$) within the charm dimuon template. This contribution has been estimated using Eq.~\eqref{eq:cont} and is less than 0.01\%. Moreover, the contribution of prompt cross-terms, such as $C_{p\bar{c}}$ and $C_{p\bar{b}}$, to the charm template are also found to be negligible, below 0.1\%. 
The charm dimuon template will also contain a contamination from beauty dimuons and mixed-terms $C_{c\bar{b}}$ ($C_{b\bar{c}}$). The magnitude of this contribution can be estimated using the production cross sections from the simulation detailed in Sec.~\ref{sec:soverbest}. However, the NLO computations have large uncertainties, and the impact of these contributions in the template varies significantly within the uncertainty range. The suppression of these backgrounds will therefore require a dedicated selection optimization in data.  

In order to demonstrate the quality of the extracted template on single muon level based on the proposed strategy, a closure test is performed. The Drell-Yan, charm and beauty dimuon simulation are combined in a single dataset according to the expected abundance in data. The probe-muon IP distribution is extracted from this data with the single tag strategy presented above. Finally, the probe-muon IP distribution from the tagging is compared with the true distribution from the charm that was simulated initially. The procedure is repeated as before for two different assumptions on the charm and beauty relative contributions. First, a charm-to-beauty production ratio of 0.5, then a ratio of 2 is also considered. 

\begin{figure}[tb!]
\centering
  \includegraphics[width=0.49\textwidth]{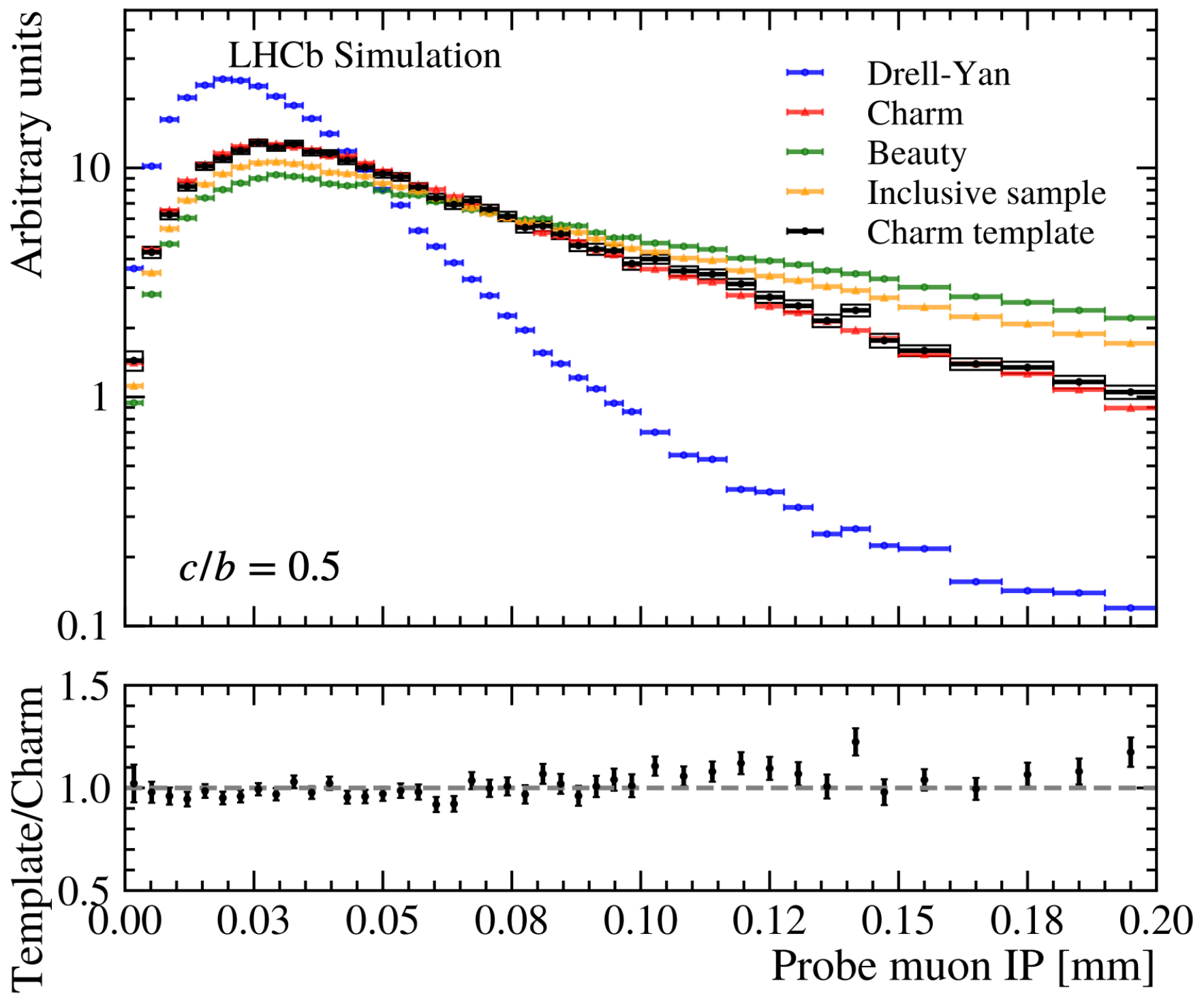}
  \includegraphics[width=0.49\textwidth]{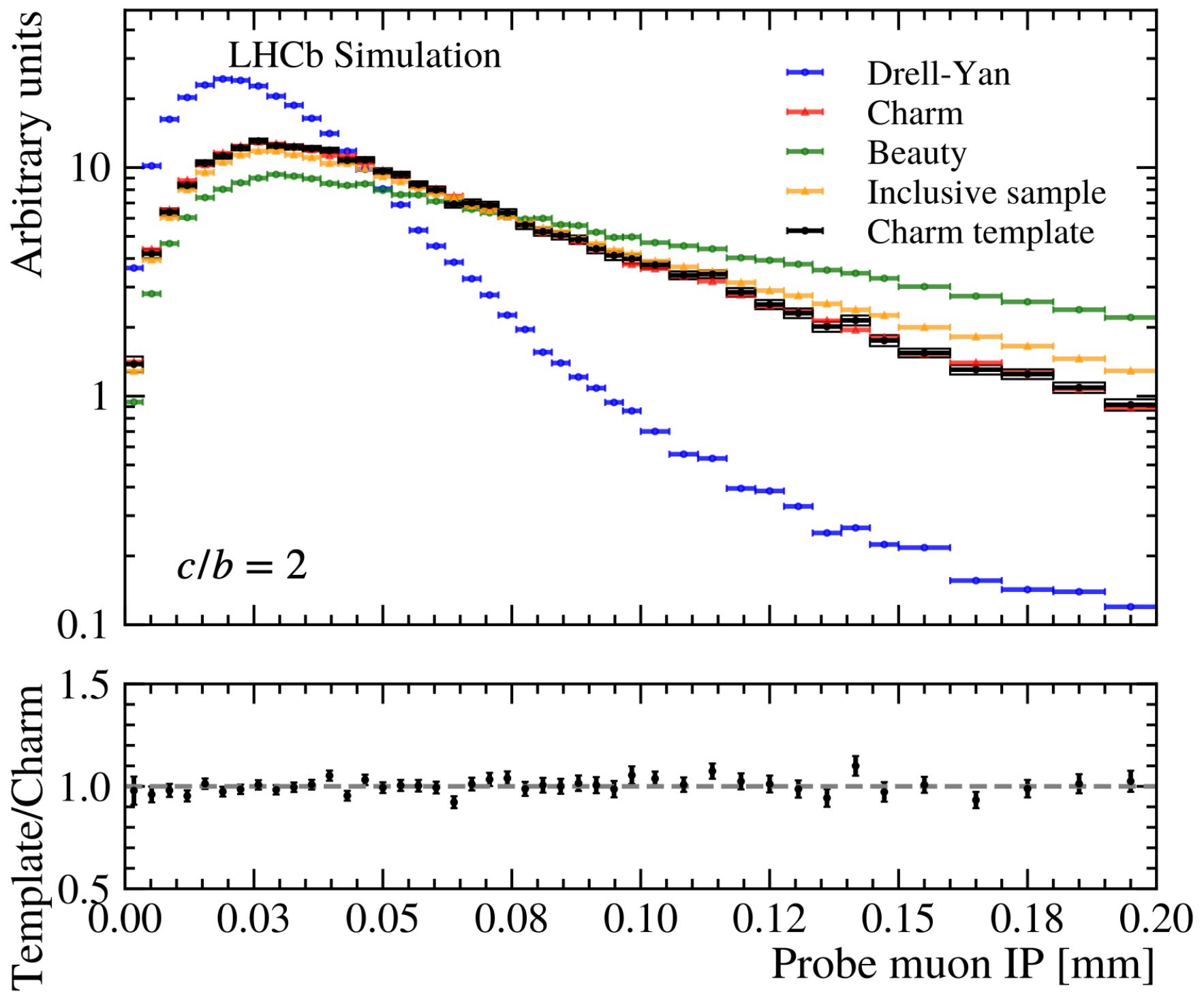}
\caption{IP distribution of the probe muon in the inclusive sample (orange), before any selections, and the final charm template after applying the single-tagging strategy (black) with boxes indicating the size of the statistical uncertainties. For comparison, the distribution of the Drell-Yan (blue), beauty (green), and charm (red) components of the inclusive sample are shown without statistical uncertainties. At the bottom, the ratio between the final template and the charm distribution. The two panels correspond to different fractions of beauty and charm ($c/b$) in the inclusive sample. }
\label{fig:closure}
\end{figure}

The results of the closure test for the probe-muon IP distribution are shown in Fig.~\ref{fig:closure} for both cases. The IP distribution of the total sample, the individual contributions as well as the distribution extracted from the single-tag algorithm are shown. In order to illustrate the quality of the extracted template, the ratio of the normalized distribution for the template and the true charm distribution is displayed below the main figure. The resulting biases observed in the ratio are weak. For the lowest IP muon values below 0.06 mm, all bins of the ratio are consistent with unity. Whereas the template ratio is consistent with unity for the larger charm over beauty cross-section case, an impact of the beauty contamination in the case of the larger initial beauty contribution can be observed.

\section{Summary and outlook}
\label{sec:out}
The small signal over background in prompt dilepton measurements below 20 \gevcc at colliders is estimated based on state-of-the-art simulations in the case of Drell-Yan production in the \lhcb acceptance at masses starting from 2.9 \gevcc. The complication arising from the scarce knowledge of semileptonic decays of charm baryons is illustrated based on variables related to the displacement of the tracks from the primary vertex. 
The presented $\mathtt{SemiCharmTag}$ tool allows to tag semileptonic decays of charm based on other charged tracks originating from the charm decay vertex. This tagging approach can be applied either on one of the two leptons (single-tag) or on both leptons (double-tag).

The double-tag tool is shown to improve the signal over background at the lowest invariant masses $2.9-5$ \gevcc by a factor $\sim 4$  at a signal efficiency of  81\%.

The single-tag is shown to provide distributions of single muon kinematics from semileptonic decays of charm that can be used for dimuon template construction. A tagging efficiency of 21.4\%, suppressing the fraction of Drell-Yan  muons by a factor of 20, is achieved based on full simulation. This method presents a data-driven approach to handle the missing knowledge of production fractions of various charm hadrons at hadron colliders and the missing information on semileptonic decays. 

These methods are ingredients to render a Drell-Yan, thermal/preequilibrium dilepton measurement down to low masses at colliders feasible despite the large background of semileptonic charm decays. 

\section*{Acknowledgements}
%
%
\noindent 
We thank the LHCb Simulation Project for the support in producing the simulated
samples used in this work. We are also grateful to the Real-Time Analysis (RTA)
Project and Data Processing and Analysis (DPA) Project for their valuable contributions
throughout the project.  A. Lightbody, I. Corredoira and Michael Winn are supported by the Agence Nationale de la recherche Grant AccessEmergence (No. ANR-23-CE31-0013). Carolina Arata and Michael Winn are supported by the université Paris-Saclay grant AccessEmergence.

\bibliographystyle{LHCb}

\bibliography{main}

@article{Ethier:2020way,
    author = "Ethier, Jacob J. and Nocera, Emanuele R.",
    title = "{Parton Distributions in Nucleons and Nuclei}",
    eprint = "2001.07722",
    archivePrefix = "arXiv",
    primaryClass = "hep-ph",
    reportNumber = "Nikhef/2020-003",
    doi = "10.1146/annurev-nucl-011720-042725",
    journal = "Ann. Rev. Nucl. Part. Sci.",
    volume = "70",
    pages = "43--76",
    year = "2020"
}

@lhcbreport{LHCb-TDR-013,
      collaboration = "LHCb collaboration",
      title         = "{LHCb VELO Upgrade Technical Design Report}",
      institution   = "CERN",
      address       = "Geneva",
      number        = "{CERN-LHCC-2013-021}",
      year          = "2013",
}

@article{Klasen:2023uqj,
    author = "Klasen, M. and Paukkunen, H.",
    title = "{Nuclear PDFs After the First Decade of LHC Data}",
    eprint = "2311.00450",
    archivePrefix = "arXiv",
    primaryClass = "hep-ph",
    reportNumber = "MS-TP-23-45",
    doi = "10.1146/annurev-nucl-102122-022747",
    journal = "Ann. Rev. Nucl. Part. Sci.",
    volume = "74",
    pages = "49--87",
    year = "2024"
}

@article{Bailhache:2025kwa,
    author = {Bailhache, R. and Appelsh{\"a}user, H.},
    title = "{Dileptons at Colliders as Probes of the Quark{\textendash}Gluon Plasma}",
    eprint = "2512.10597",
    archivePrefix = "arXiv",
    primaryClass = "nucl-ex",
    doi = "10.1146/annurev-nucl-121423-100858",
    journal = "Ann. Rev. Nucl. Part. Sci.",
    volume = "75",
    number = "1",
    pages = "463--486",
    year = "2025"
}

@article{Salabura:2020tou,
    author = "Salabura, Piotr and Stroth, Joachim",
    title = "{Dilepton radiation from strongly interacting systems}",
    eprint = "2005.14589",
    archivePrefix = "arXiv",
    primaryClass = "nucl-ex",
    reportNumber = "https://www.sciencedirect.com/science/article/abs/pii/S0146641021000235",
    doi = "10.1016/j.ppnp.2021.103869",
    journal = "Prog. Part. Nucl. Phys.",
    volume = "120",
    pages = "103869",
    year = "2021"
}

@article{Coquet:2021lca,
    author = "Coquet, Maurice and Du, Xiaojian and Ollitrault, Jean-Yves and Schlichting, Soeren and Winn, Michael",
    title = "{Intermediate mass dileptons as pre-equilibrium probes in heavy ion collisions}",
    eprint = "2104.07622",
    archivePrefix = "arXiv",
    primaryClass = "nucl-th",
    doi = "10.1016/j.physletb.2021.136626",
    journal = "Phys. Lett. B",
    volume = "821",
    pages = "136626",
    year = "2021"
}

@article{NA60:2008ctj,
    author = "Arnaldi, R. and others",
    collaboration = "NA60",
    title = "{NA60 results on thermal dimuons}",
    eprint = "0812.3053",
    archivePrefix = "arXiv",
    primaryClass = "nucl-ex",
    doi = "10.1140/epjc/s10052-009-0878-5",
    journal = "Eur. Phys. J. C",
    volume = "61",
    pages = "711--720",
    year = "2009"
}

@article{ALEPH:2005ab,
    author = "Schael, S. and others",
    collaboration = "ALEPH, DELPHI, L3, OPAL, SLD, LEP Electroweak Working Group, SLD Electroweak Group, SLD Heavy Flavour Group",
    title = "{Precision electroweak measurements on the $Z$ resonance}",
    eprint = "hep-ex/0509008",
    archivePrefix = "arXiv",
    reportNumber = "SLAC-R-774",
    doi = "10.1016/j.physrep.2005.12.006",
    journal = "Phys. Rept.",
    volume = "427",
    pages = "257--454",
    year = "2006"
}

@article{Alwall:2014hca,
    author = "Alwall, J. and Frederix, R. and Frixione, S. and Hirschi, V. and Maltoni, F. and Mattelaer, O. and Shao, H. -S. and Stelzer, T. and Torrielli, P. and Zaro, M.",
    title = "{The automated computation of tree-level and next-to-leading order differential cross sections, and their matching to parton shower simulations}",
    eprint = "1405.0301",
    archivePrefix = "arXiv",
    primaryClass = "hep-ph",
    reportNumber = "CERN-PH-TH-2014-064, CP3-14-18, LPN14-066, MCNET-14-09, ZU-TH-14-14",
    doi = "10.1007/JHEP07(2014)079",
    journal = "JHEP",
    volume = "07",
    pages = "079",
    year = "2014"
}

@article{Frederix:2018nkq,
    author = "Frederix, R. and Frixione, S. and Hirschi, V. and Pagani, D. and Shao, H. -S. and Zaro, M.",
    title = "{The automation of next-to-leading order electroweak calculations}",
    eprint = "1804.10017",
    archivePrefix = "arXiv",
    primaryClass = "hep-ph",
    reportNumber = "Nikhef/2018-015, TUM-HEP-1138/18, NIKHEF-2018-015, TUM-HEP-1138-18",
    doi = "10.1007/JHEP11(2021)085",
    journal = "JHEP",
    volume = "07",
    pages = "185",
    year = "2018",
    note = "[Erratum: JHEP 11, 085 (2021)]"
}

@article{PDG2024,
     author    = "Navas, S. and others",
    collaboration = "Particle Data Group",
     title     = "{\href{http://pdg.lbl.gov/}{Review of particle physics}}",
     journal   = "Phys. Rev.",
     year = {2024},
     number = {8},
     volume      = "D110",
     pages     = "030001",
     doi = "10.1103/PhysRevD.110.030001"
}

@article{Li:2025nzx,
    author = "Li, Pei-Rong and Lyu, Xiao-Rui and Zheng, Yangheng",
    title = "{Experimental overview on the charmed baryon decays}",
    eprint = "2509.19141",
    archivePrefix = "arXiv",
    primaryClass = "hep-ex",
    doi = "10.1088/1674-1137/ae1187",
    journal = "Chin. Phys. C",
    volume = "50",
    pages = "022002",
    year = "2026"
}

@article{Sjostrand:2007gs,
    author = "Sjostrand, Torbjorn and Mrenna, Stephen and Skands, Peter Z.",
    title = "{A Brief Introduction to PYTHIA 8.1}",
    eprint = "0710.3820",
    archivePrefix = "arXiv",
    primaryClass = "hep-ph",
    reportNumber = "CERN-LCGAPP-2007-04, LU-TP-07-28, FERMILAB-PUB-07-512-CD-T",
    doi = "10.1016/j.cpc.2008.01.036",
    journal = "Comput. Phys. Commun.",
    volume = "178",
    pages = "852--867",
    year = "2008"
}

@techreport{Belyaev:1322400,
      author        = "Belyaev, I and Brambach, T and Brook, N H and Gauvin, N
                       and Corti, G and Harrison, K and Harrison, P F and He, J
                       and Jones, C R and Lieng, M and Manca, G and Miglioranzi, S
                       and Robbe, P and Vagnoni, V and Whitehead, M and Wishahi,
                       J",
      title         = "{Handling of the generation of primary events in Gauss,
                       the LHCb simulation framework}",
      institution   = "CERN",
      reportNumber  = "LHCb-PROC-2011-005, CERN-LHCb-PROC-2011-005",
      address       = "Geneva",
      year          = "2011",
      note          = "LHCb-PROC-2011-005, CERN-LHCb-PROC-2011-005",
      url           = "https://cds.cern.ch/record/1322400",
}

@Article{Lange:2001uf,
     author    = "Lange, D. J.",
     title     = "{The EvtGen particle decay simulation package}",
     journal   = "Nucl. Instrum. Meth.",
     volume    = "A462",
     year      = "2001",
     pages     = "152-155",
     doi       = "10.1016/S0168-9002(01)00089-4",
}

@article{Golonka:2005pn,
      author         = "Golonka, Piotr and Was, Zbigniew",
      title          = "{PHOTOS Monte Carlo: A precision tool for QED
                  corrections in $Z$ and $W$ decays}",
      journal        = "Eur. Phys. J.",
      volume         = "C45",
      pages          = "97-107",
      doi            = "10.1140/epjc/s2005-02396-4",
      year           = "2006",
      eprint         = "hep-ph/0506026",
      archivePrefix  = "arXiv",
      primaryClass   = "hep-ph",
      reportNumber   = "IFJPAN-V-05-01, CERN-PH-TH-2005-091",
}

@Article{Agostinelli:2002hh,
     author    = "Agostinelli, S. and others",
 collaboration = "Geant4 collaboration",
     title     = "{Geant4: A simulation toolkit}",
     journal   = "Nucl. Instrum. Meth.",
     volume    = "A506",
     year      = "2003",
     pages     = "250",
     doi       = "10.1016/S0168-9002(03)01368-8",
}

@article{Allison:2006ve,
      author         = "Allison, John and Amako, K. and Apostolakis, J. and
                        Araujo, H. and Dubois, P.A. and others",
 collaboration = "Geant4 collaboration",
      title          = "{Geant4 developments and applications}",
      journal        = "IEEE Trans.Nucl.Sci.",
      volume         = "53",
      pages          = "270",
      doi            = "10.1109/TNS.2006.869826",
      year           = "2006",
      reportNumber   = "SLAC-PUB-11870",
}

@article{LHCb-PROC-2011-006,
  author="Clemencic, M and others",
  title="{The \lhcb simulation application, Gauss: Design, evolution and experience}",
  journal="J. Phys. Conf. Ser.",
  volume={331},
  pages={032023},
  doi={10.1088/1742-6596/331/3/032023},
  year={2011},
}

@article{MarcoClemencic_2010,
doi = {10.1088/1742-6596/219/4/042006},
url = {https://doi.org/10.1088/1742-6596/219/4/042006},
year = {2010},
month = {apr},
publisher = {},
volume = {219},
number = {4},
pages = {042006},
author = {Marco Clemencic and Hubert Degaudenzi and Pere Mato and Sebastien Binet and Wim Lavrijsen and Charles Leggett and Ivan Belyaev},
title = {Recent developments in the {LHC}b software framework gaudi},
journal = {Journal of Physics: Conference Series}
}

@article{BARRAND200145,
title = {GAUDI — A software architecture and framework for building HEP data processing applications},
journal = {Computer Physics Communications},
volume = {140},
number = {1},
pages = {45-55},
year = {2001},
note = {CHEP2000},
issn = {0010-4655},
doi = {https://doi.org/10.1016/S0010-4655(01)00254-5},
url = {https://www.sciencedirect.com/science/article/pii/S0010465501002545},
author = {G. Barrand and I. Belyaev and P. Binko and M. Cattaneo and R. Chytracek and G. Corti and M. Frank and G. Gracia and J. Harvey and E.van Herwijnen and P. Maley and P. Mato and S. Probst and F. Ranjard}
}

@article{Gelis:2010nm,
    author = "Gelis, Francois and Iancu, Edmond and Jalilian-Marian, Jamal and Venugopalan, Raju",
    title = "{The Color Glass Condensate}",
    eprint = "1002.0333",
    archivePrefix = "arXiv",
    primaryClass = "hep-ph",
    doi = "10.1146/annurev.nucl.010909.083629",
    journal = "Ann. Rev. Nucl. Part. Sci.",
    volume = "60",
    pages = "463--489",
    year = "2010"
}

@article{Boussarie:2023izj,
    author = "Boussarie, Renaud and others",
    title = "{TMD Handbook}",
    eprint = "2304.03302",
    archivePrefix = "arXiv",
    primaryClass = "hep-ph",
    reportNumber = "JLAB-THY-23-3780, LA-UR-21-20798, MIT-CTP/5386",
    month = "4",
    year = "2023"
}

@article{Cacciari:2012ny,
    author = "Cacciari, Matteo and Frixione, Stefano and Houdeau, Nicolas and Mangano, Michelangelo L. and Nason, Paolo and Ridolfi, Giovanni",
    title = "{Theoretical predictions for charm and bottom production at the LHC}",
    eprint = "1205.6344",
    archivePrefix = "arXiv",
    primaryClass = "hep-ph",
    reportNumber = "CERN-PH-TH-2011-227",
    doi = "10.1007/JHEP10(2012)137",
    journal = "JHEP",
    volume = "10",
    pages = "137",
    year = "2012"
}

@article{Cacciari:2015fta,
    author = "Cacciari, Matteo and Mangano, Michelangelo L. and Nason, Paolo",
    title = "{Gluon PDF constraints from the ratio of forward heavy-quark production at the LHC at $\sqrt{s}=7$ and 13 TeV}",
    eprint = "1507.06197",
    archivePrefix = "arXiv",
    primaryClass = "hep-ph",
    reportNumber = "CERN-PH-TH-2015-171",
    doi = "10.1140/epjc/s10052-015-3814-x",
    journal = "Eur. Phys. J. C",
    volume = "75",
    number = "12",
    pages = "610",
    year = "2015"
}

@article{ALICE:2019wqv,
    author = "Acharya, Shreyasi and others",
    collaboration = "ALICE",
    title = "{Measurement of charged jet cross section in $pp$ collisions at ${\sqrt{s}=5.02}$ TeV}",
    eprint = "1905.02536",
    archivePrefix = "arXiv",
    primaryClass = "nucl-ex",
    reportNumber = "CERN-EP-2019-070",
    doi = "10.1103/PhysRevD.100.092004",
    journal = "Phys. Rev. D",
    volume = "100",
    number = "9",
    pages = "092004",
    year = "2019"
}

@article{Sjostrand:2014zea,
    author = {Sj{\"o}strand, Torbj{\"o}rn and Ask, Stefan and Christiansen, Jesper R. and Corke, Richard and Desai, Nishita and Ilten, Philip and Mrenna, Stephen and Prestel, Stefan and Rasmussen, Christine O. and Skands, Peter Z.},
    title = "{An introduction to PYTHIA 8.2}",
    eprint = "1410.3012",
    archivePrefix = "arXiv",
    primaryClass = "hep-ph",
    reportNumber = "LU-TP-14-36, MCNET-14-22, CERN-PH-TH-2014-190, FERMILAB-PUB-14-316-CD, DESY-14-178, SLAC-PUB-16122",
    doi = "10.1016/j.cpc.2015.01.024",
    journal = "Comput. Phys. Commun.",
    volume = "191",
    pages = "159--177",
    year = "2015"
}

@article{Apolinario:2022vzg,
    author = "Apolin{\'a}rio, Liliana and Lee, Yen-Jie and Winn, Michael",
    title = "{Heavy quarks and jets as probes of the QGP}",
    eprint = "2203.16352",
    archivePrefix = "arXiv",
    primaryClass = "hep-ph",
    doi = "10.1016/j.ppnp.2022.103990",
    journal = "Prog. Part. Nucl. Phys.",
    volume = "127",
    pages = "103990",
    year = "2022"
}

@article{ATLAS:2014ape,
    author = "Aad, Georges and others",
    collaboration = "ATLAS",
    title = "{Measurement of the low-mass Drell-Yan differential cross section at $\sqrt{s}$ = 7 TeV using the ATLAS detector}",
    eprint = "1404.1212",
    archivePrefix = "arXiv",
    primaryClass = "hep-ex",
    reportNumber = "CERN-PH-EP-2014-020",
    doi = "10.1007/JHEP06(2014)112",
    journal = "JHEP",
    volume = "06",
    pages = "112",
    year = "2014"
}

@article{ALICE:2023sgl,
    author = "Acharya, Shreyasi and others",
    collaboration = "ALICE",
    title = "{Charm production and fragmentation fractions at midrapidity in pp collisions at $ \sqrt{\textrm{s}} $ = 13 TeV}",
    eprint = "2308.04877",
    archivePrefix = "arXiv",
    primaryClass = "hep-ex",
    reportNumber = "CERN-EP-2023-162",
    doi = "10.1007/JHEP12(2023)086",
    journal = "JHEP",
    volume = "12",
    pages = "086",
    year = "2023"
}

@article{CMS:2021ynu,
    author = "Sirunyan, Albert M and others",
    collaboration = "CMS",
    title = "{Study of Drell-Yan dimuon production in proton-lead collisions at $\sqrt{s_\mathrm{NN}} =$ 8.16 TeV}",
    eprint = "2102.13648",
    archivePrefix = "arXiv",
    primaryClass = "hep-ex",
    reportNumber = "CMS-HIN-18-003, CERN-EP-2021-028",
    doi = "10.1007/JHEP05(2021)182",
    journal = "JHEP",
    volume = "05",
    pages = "182",
    year = "2021"
}

@article{Chen:2016btl,
    author = "Chen, Tianqi and Guestrin, Carlos",
    title = "{XGBoost: A Scalable Tree Boosting System}",
    eprint = "1603.02754",
    archivePrefix = "arXiv",
    primaryClass = "cs.LG",
    doi = "10.1145/2939672.2939785",
    month = "3",
    year = "2016"
}

@article{Bellm:2015jjp,
    author = "Bellm, Johannes and others",
    title = "{Herwig 7.0/Herwig++ 3.0 release note}",
    eprint = "1512.01178",
    archivePrefix = "arXiv",
    primaryClass = "hep-ph",
    reportNumber = "CERN-PH-TH-2015-289, MAN-HEP-2015-15, IFJPAN-IV-2015-13, KA-TP-18-2015, DCPT-15-142, MCNET-15-28, IPPP-15-71, HERWIG-2015-01",
    doi = "10.1140/epjc/s10052-016-4018-8",
    journal = "Eur. Phys. J. C",
    volume = "76",
    number = "4",
    pages = "196",
    year = "2016"
}

@article{NNNPDF30,
   title={Parton distributions for the {LHC} Run2},
   volume={2015},
   ISSN={1029-8479},
   url={http://dx.doi.org/10.1007/JHEP04(2015)040},
   DOI={10.1007/jhep04(2015)040},
   number={4},
   journal={Journal of High Energy Physics},
   publisher={Springer Science and Business Media LLC},
   author={Ball, Richard D. and Bertone, Valerio and Carrazza, Stefano and Deans, Christopher S. and Del Debbio, Luigi and Forte, Stefano and Guffanti, Alberto and Hartland, Nathan P. and Latorre, José I. and Rojo, Juan and Ubiali, Maria},
   year={2015},
   month=apr }

@article{Skands:2014pea,
    author = "Skands, Peter and Carrazza, Stefano and Rojo, Juan",
    title = "{Tuning PYTHIA 8.1: the Monash 2013 Tune}",
    eprint = "1404.5630",
    archivePrefix = "arXiv",
    primaryClass = "hep-ph",
    reportNumber = "CERN-PH-TH-2014-069, MCNET-14-08, OUTP-14-05P",
    doi = "10.1140/epjc/s10052-014-3024-y",
    journal = "Eur. Phys. J. C",
    volume = "74",
    number = "8",
    pages = "3024",
    year = "2014"
}

@article{LHCb:2017vec,
    author = "Aaij, Roel and others",
    collaboration = "LHCb",
    title = "{Measurement of the $B^{\pm}$ production cross-section in pp collisions at $\sqrt{s} =$ 7 and 13 TeV}",
    eprint = "1710.04921",
    archivePrefix = "arXiv",
    primaryClass = "hep-ex",
    reportNumber = "CERN-EP-2017-254, LHCB-PAPER-2017-037",
    doi = "10.1007/JHEP12(2017)026",
    journal = "JHEP",
    volume = "12",
    pages = "026",
    year = "2017"
}

@article{LHCb:2013xam,
    author = "Aaij, R and others",
    collaboration = "LHCb",
    title = "{Prompt charm production in pp collisions at sqrt(s)=7 TeV}",
    eprint = "1302.2864",
    archivePrefix = "arXiv",
    primaryClass = "hep-ex",
    reportNumber = "LHCB-PAPER-2012-041, CERN-PH-EP-2013-009",
    doi = "10.1016/j.nuclphysb.2013.02.010",
    journal = "Nucl. Phys. B",
    volume = "871",
    pages = "1--20",
    year = "2013"
}

@article{ALICE:2021dhb,
    author = "Acharya, Shreyasi and others",
    collaboration = "ALICE",
    title = "{Charm-quark fragmentation fractions and production cross section at midrapidity in pp collisions at the LHC}",
    eprint = "2105.06335",
    archivePrefix = "arXiv",
    primaryClass = "nucl-ex",
    reportNumber = "CERN-EP-2021-088",
    doi = "10.1103/PhysRevD.105.L011103",
    journal = "Phys. Rev. D",
    volume = "105",
    number = "1",
    pages = "L011103",
    year = "2022"
}

@article{LHCb:2017rmj,
    author = "Aaij, Roel and others",
    collaboration = "LHCb",
    title = "{Measurement of the $B^0_s\to\mu^+\mu^-$ branching fraction and effective lifetime and search for $B^0\to\mu^+\mu^-$ decays}",
    eprint = "1703.05747",
    archivePrefix = "arXiv",
    primaryClass = "hep-ex",
    reportNumber = "CERN-EP-2017-041, LHCB-PAPER-2017-001",
    doi = "10.1103/PhysRevLett.118.191801",
    journal = "Phys. Rev. Lett.",
    volume = "118",
    number = "19",
    pages = "191801",
    year = "2017"
}

@article{LHCb:2015swx,
    author = "Aaij, Roel and others",
    collaboration = "LHCb",
    title = "{Measurements of prompt charm production cross-sections in $pp$ collisions at $ \sqrt{s}=$ 13 TeV}",
    eprint = "1510.01707",
    archivePrefix = "arXiv",
    primaryClass = "hep-ex",
    reportNumber = "LHCB-PAPER-2015-041, CERN-PH-EP-2015-272",
    doi = "10.1007/JHEP03(2016)159",
    journal = "JHEP",
    volume = "03",
    pages = "159",
    year = "2016",
    note = "[Erratum: JHEP 09, 013 (2016), Erratum: JHEP 05, 074 (2017)]"
}

@article{ALICE:2018fvj,
    author = "Acharya, Shreyasi and others",
    collaboration = "ALICE",
    title = "{Dielectron production in proton-proton collisions at $ \sqrt{s}=7 $ TeV}",
    eprint = "1805.04391",
    archivePrefix = "arXiv",
    primaryClass = "hep-ex",
    reportNumber = "CERN-EP-2018-102",
    doi = "10.1007/JHEP09(2018)064",
    journal = "JHEP",
    volume = "09",
    pages = "064",
    year = "2018"
}

@article{ALICE:2023jef,
    author = "Acharya, Shreyasi and others",
    collaboration = "ALICE",
    title = "{Dielectron production in central Pb-Pb collisions at sNN=5.02TeV}",
    eprint = "2308.16704",
    archivePrefix = "arXiv",
    primaryClass = "nucl-ex",
    reportNumber = "CERN-EP-2023-194",
    doi = "10.1103/xl6m-vbqk",
    journal = "Phys. Rev. C",
    volume = "112",
    number = "5",
    pages = "054906",
    year = "2025"
}

@article{LHCb:2021dlw,
    author = "Aaij, Roel and others",
    collaboration = "LHCb",
    title = "{Identification of charm jets at LHCb}",
    eprint = "2112.08435",
    archivePrefix = "arXiv",
    primaryClass = "hep-ex",
    reportNumber = "LHCb-DP-2021-006",
    doi = "10.1088/1748-0221/17/02/P02028",
    journal = "JINST",
    volume = "17",
    number = "02",
    pages = "P02028",
    year = "2022"
}

@article{Anderlini:2020ucv,
    author = "Anderlini, Lucio and others",
    title = "{Muon identification for LHCb Run 3}",
    eprint = "2008.01579",
    archivePrefix = "arXiv",
    primaryClass = "hep-ex",
    reportNumber = "LHCb-DP-2020-002",
    doi = "10.1088/1748-0221/15/12/T12005",
    journal = "JINST",
    volume = "15",
    number = "12",
    pages = "T12005",
    year = "2020"
}

@misc{web:davinci,
  title = "{DaVinci application GitLab repository}",
  collaboration = "LHCb",
  url = {https://gitlab.cern.ch/lhcb/DaVinci/},
  note = "{https://gitlab.cern.ch/lhcb/DaVinci/}"
}

@article{CDF_Jpsi,
  title = "{Measurement of the average lifetime of B hadrons produced in $p\bar{p}$ collisions at $ \sqrt{s}=$ 1.8 TeV}",
  author = {Abe, F. and others},
  journal = {Phys. Rev. Lett.},
  collaboration = {CDF collaboration},
  volume = {71},
  issue = {21},
  pages = {3421--3426},
  numpages = {0},
  year = {1993},
  month = {Nov},
  publisher = {American Physical Society},
  doi = {10.1103/PhysRevLett.71.3421},
  url = {https://link.aps.org/doi/10.1103/PhysRevLett.71.3421}
}

@article{Dark_photons_LHCb,
    author = "Aaij, Roel and others",
    collaboration = "LHCb",
    title = "{Search for Dark Photons Produced in 13 TeV $pp$ Collisions}",
    eprint = "1710.02867",
    archivePrefix = "arXiv",
    primaryClass = "hep-ex",
    reportNumber = "LHCB-PAPER-2017-038, CERN-EP-2017-248",
    doi = "10.1103/PhysRevLett.120.061801",
    journal = "Phys. Rev. Lett.",
    volume = "120",
    number = "6",
    pages = "061801",
    year = "2018"
}
 
\newpage

\end{document}